\documentclass[preprint,12pt]{aastex}

\newcommand{\jhks}{\ensuremath{JHK_s}}
\newcommand{\spitzer}{\textit{Spitzer}}
\newcommand{\ak}{\ensuremath{A_{K_s}}}
\newcommand{\chisq}{\ensuremath{\chi^2}}

\newcommand{\alak}{\ensuremath{A_\lambda/A_{K_s}}}

\begin{document}
\author{Nicholas L.\ Chapman\altaffilmark{1,2},
Lee G.\ Mundy\altaffilmark{1},
Shih-Ping Lai\altaffilmark{3},
Neal J.\ Evans II\altaffilmark{4}}

\altaffiltext{1}{Department of Astronomy, University of Maryland, College Park,
MD 20742}
\altaffiltext{2}{Jet Propulsion Laboratory, California Institute of Technology,
4800 Oak Grove Drive, MS 301-429, Pasadena, CA 91109; 
Nicholas.L.Chapman@jpl.nasa.gov}
\altaffiltext{3}{Institute of Astronomy and Department of Physics,
National Tsing Hua University, Hsinchu 30043, Taiwan}
\altaffiltext{4}{Department of Astronomy, University of Texas at Austin,
1 University Station C1400, Austin, TX 78712}

\title{\label{sec:clouds} The Mid-Infrared Extinction Law in the Ophiuchus,
Perseus, and Serpens Molecular Clouds}

\begin{abstract}

We compute the mid-infrared extinction law from $3.6-24\:\mu$m in three
molecular clouds: Ophiuchus, Perseus, and Serpens, by combining data from the
``Cores to Disks'' \spitzer{} Legacy Science program with deep \jhks{} imaging.
Using a new technique, we are able to calculate the line-of-sight extinction law
towards each background star in our fields. With these line-of-sight
measurements, we create, for the first time, maps of the \chisq{} deviation of
the data from two extinction law models.  Because our \chisq{} maps have the
same spatial resolution as our extinction maps, we can directly observe the
changing  extinction law as a function of the total column density.  In the
\spitzer{} IRAC bands, $3.6-8\:\mu$m, we see evidence for grain growth.  Below
$A_{K_s} = 0.5$, our extinction law is well-fit by the \citet{weingartner01}
$R_V = 3.1$ diffuse interstellar medium dust model.  As the extinction
increases, our law gradually flattens, and for $A_{K_s} \ge 1$, the data are
more consistent with the Weingartner \& Draine $R_V = 5.5$ model that uses
larger maximum dust grain sizes.  At $24\:\mu$m, our extinction law is
$2-4\times$ higher than the values predicted by theoretical dust models, but is
more consistent with the observational results of \citet{flaherty07}.  Lastly,
from our \chisq{} maps we identify a region in Perseus where the IRAC extinction
law is anomalously high considering its column density.  A steeper near-infrared
extinction law than the one we have assumed may partially explain the IRAC
extinction law in this region.

\end{abstract}


\section{ Introduction }

Dust is a ubiquitous component of the interstellar medium. It is important to
understand the dust because it attenuates, or extincts, the light from
background objects seen through the dust. The amount of extinction varies with
both wavelength and position. Because of the constancy of the gas-to-dust ratio,
the quantity of dust gives you a direct measure of the cloud's mass
\citep{bohlin78}. Note, however that the \citet{bohlin78} equation does depend
on the $E(B-V)$ color excess, which can potentially vary depending on
interstellar medium properties such as metallicity.  The dust extinction must
also be accounted for when computing the luminosities for protostars within
molecular clouds. The wavelength dependence of the extinction, known as the
extinction law, directly relates to dust grain properties such as size,
composition, and the presence or lack of icy mantles on the grains.

Many authors have found that the near-infrared extinction law has the form of a
power law, $A_\lambda \propto \lambda^{-\beta}$, between $\lambda \sim 1\:\mu$m
and $\sim4\:\mu$m with $\beta = 1.6-1.8$ \citep[and references
therein]{draine03}. The extinction law in the mid-infrared wavelengths has not
been as well studied. \citet{rieke85} found the near-infrared power law extended
out to $\sim7\:\mu$m before the extinctions increased again towards the
$10\:\mu$m silicate absorption peak. More recent results have found a different
behavior. \citet{lutz99} used ISO to measure hydrogen recombination lines
towards the Galactic center and \citet{indebetouw05} and \citet{flaherty07} used
the Spitzer Space telescope to measure the mid-infrared extinction law in
several regions. These authors all find a similar extinction law in the
mid-infrared which is much flatter than the \citet{rieke85} result.

To date no one has studied the changes in the extinction law within a large
region with high spatial resolution. In this paper we use \spitzer{} and deep
ground-based \jhks{} observations to probe changes in the dust properties within
three molecular clouds: Ophiuchus, Perseus, and Serpens. We compare our results
with different dust models to investigate the relationship between dust grain
properties and total extinction (column density). In \S\,\ref{sec:obs} we
describe our observations. Then, in \S\,\ref{sec:clouddatareduce} we discuss our
data reduction method, including how we identified background stars and computed
line-of-sight (LOS) extinctions. In
\S\,\ref{sec:cloudext}-\ref{sec:dustprop-clouds} we create extinction and
\chisq{} maps and compare them to each other. We also compute the average
extinction law from $3.6-24\:\mu$m for different ranges of extinction. Finally,
we summarize our results in \S\,\ref{sec:cloud-discuss}.

\section{ \label{sec:obs} Observations}

We mapped regions within the Ophiuchus, Perseus, and Serpens molecular clouds in
eight wavebands from $1.25-24\:\mu$m. Our \spitzer{} observations
($3.6-24\:\mu$m) were part of the \spitzer{} Legacy Science program ``From
Molecular Cores to Planet-Forming Disks'' (c2d) \citep{evans03}. The \spitzer{}
observations have been previously published for Ophiuchus \citep{padgett08},
Perseus \citep{jorgensen06,rebull07}, and Serpens \citep{harvey06,harvey07a}. To
complement these data, we took \jhks{} observations at Kitt Peak National
Observatory in three observing runs from September 2004-June 2006. We used the
FLoridA Multi-object Imaging Near-ir Grism Observations Spectrometer (FLAMINGOS)
\citep{elston98} on the 4-meter telescope.

The c2d observations of Ophiuchus, Perseus, and Serpens cover approximately 11
deg$^2$ with both \spitzer's IRAC ($3.6-8\:\mu$m) and MIPS ($24\:\mu$m)
instruments. Such a large area could not feasibly be observed by FLAMINGOS.
Instead, we focused on selected regions within these clouds. Our goal was to
explore changes in the extinction law within molecular clouds. Therefore, we
chose contiguous regions with a range of extinction values, from low to high.
Furthermore, we tried to include some star-forming regions within our mapped
areas since these regions can alter the dust properties as well. In Table
\ref{tab:cloudbasic} we list the basic properties of these clouds including the
total area mapped with our \jhks{} observations.

\begin{figure}

\plotone{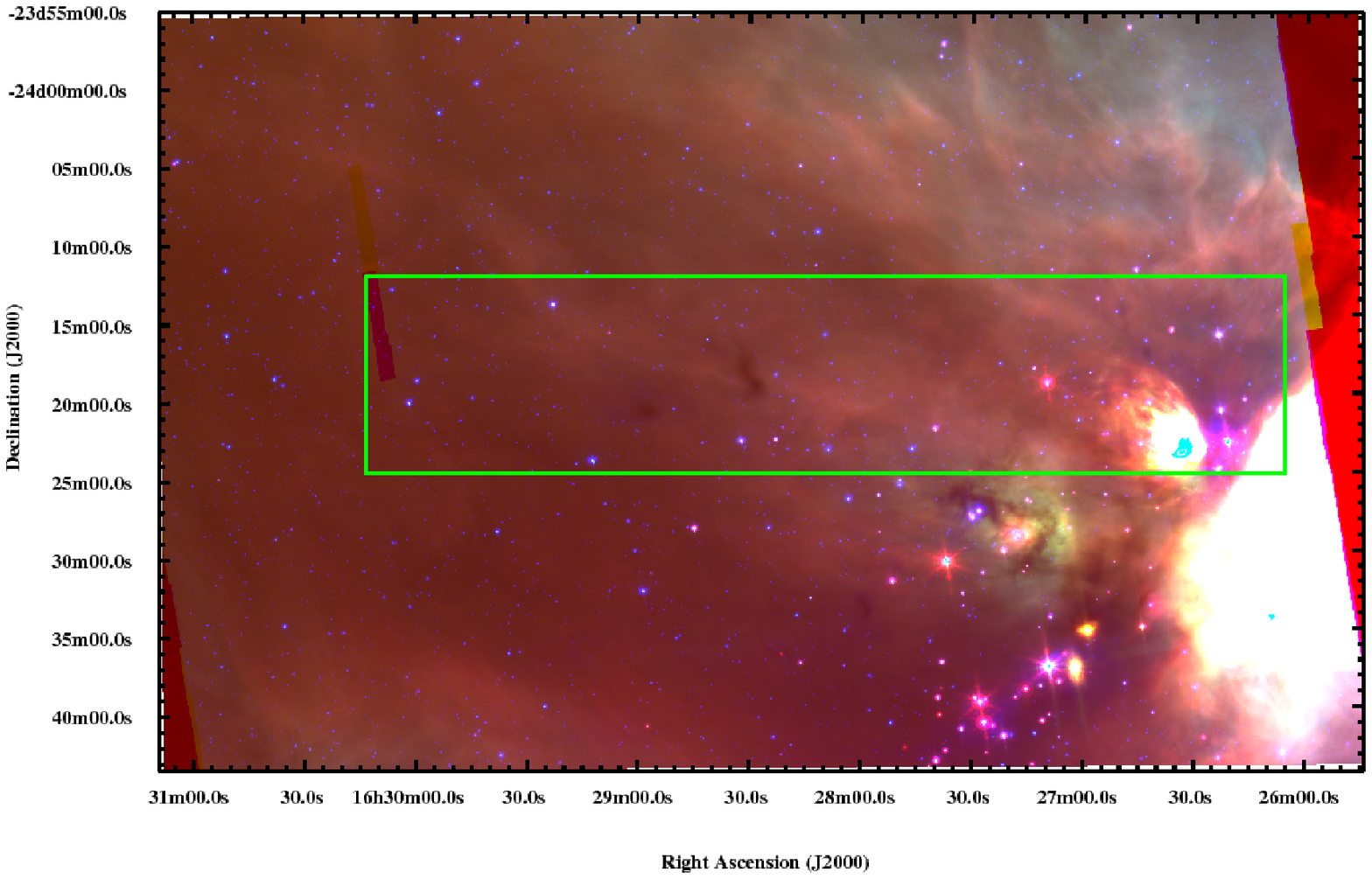}

\caption[ Color Image of Ophiuchus ]{\label{fig:ophrgb} Color image of the
Ophiuchus cloud made from the $3.6\:\mu$m (blue), $8.0\:\mu$m (green), and
$24\:\mu$m (red) channels on \spitzer. The green outline denotes the region we
observed with our \jhks{} observations. }

\end{figure}

\begin{figure}

\plotone{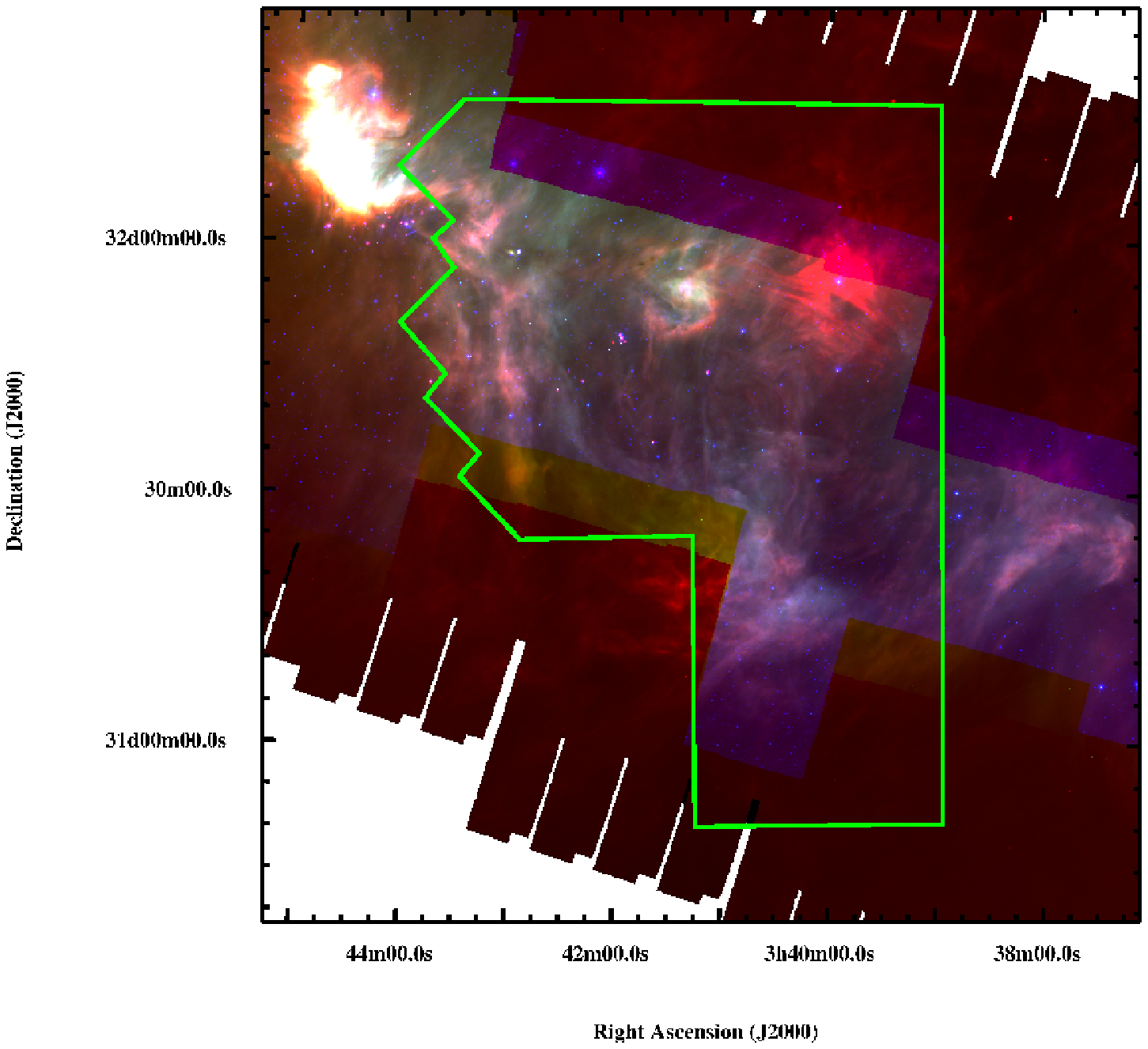}

\caption[ Color Image of Perseus ]{\label{fig:perseusrgb} Color image of the
Perseus cloud made from the $3.6\:\mu$m (blue), $8.0\:\mu$m (green), and
$24\:\mu$m (red) channels on \spitzer. The green outline denotes the region we
observed with our \jhks{} observations. }

\end{figure}

\begin{figure}

\plotone{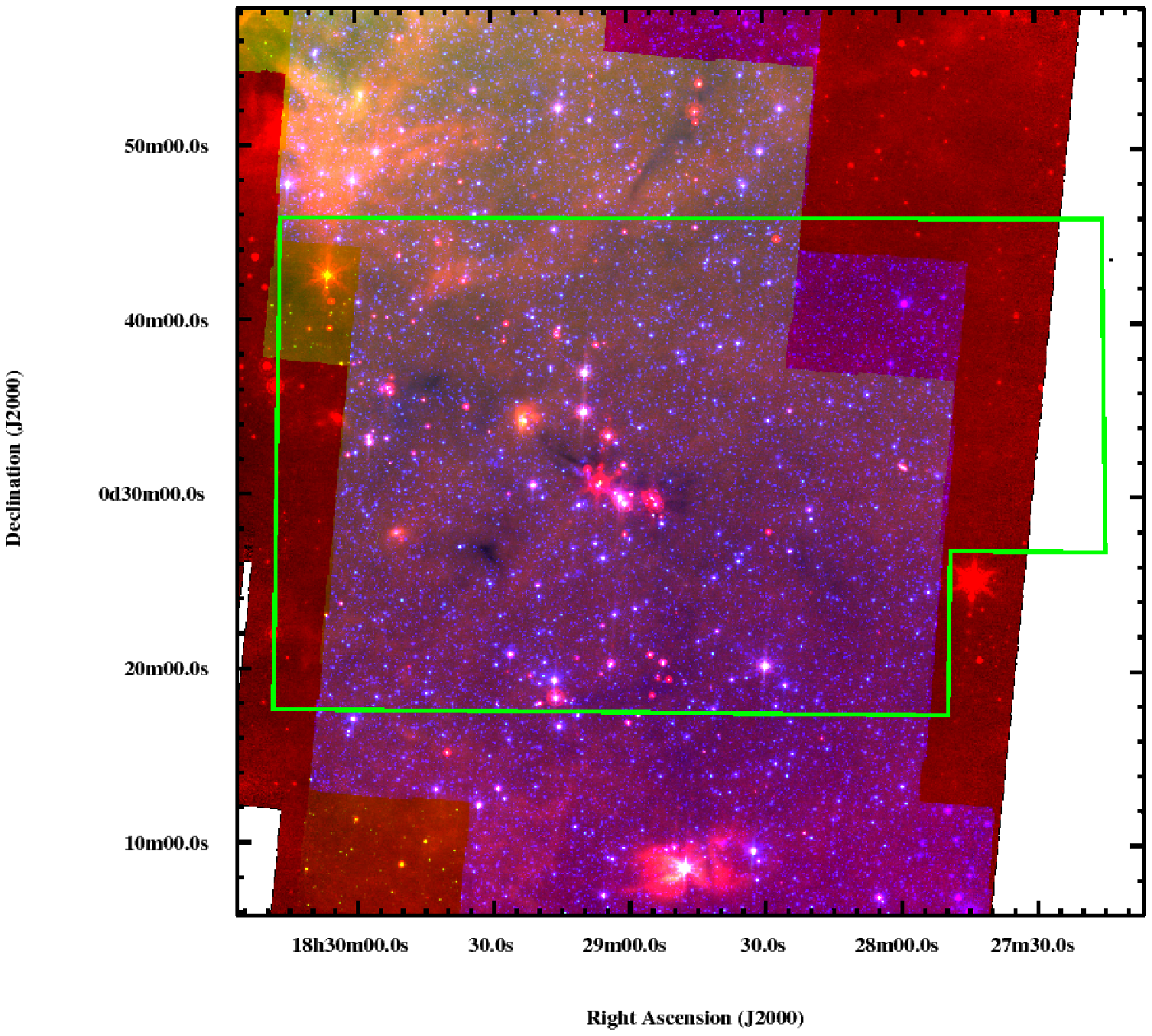}

\caption[ Color Image of Serpens ]{\label{fig:serpensrgb} Color image of the
Serpens cloud made from the $3.6\:\mu$m (blue), $8.0\:\mu$m (green), and
$24\:\mu$m (red) channels on \spitzer. The green outline denotes the region we
observed with our \jhks{} observations. }

\end{figure}

In Figures \ref{fig:ophrgb} - \ref{fig:serpensrgb} we plotted a three color
image of each region using the c2d data. The red, green, and blue emission
correspond to the 24, 8, and $3.6\:\mu$m channels on \spitzer, respectively. The
green outline denotes the area we mapped in the \jhks{} bands.

In Ophiuchus we mapped a $\sim 10\arcmin \times 60\arcmin$ region to the
northeast of L1688. The western side of our map, near L1688, contains many
young stellar objects (YSOs) and several mid-infrared dark cores are seen in
the middle and on the eastern side. The $24\:\mu$m emission is particularly
bright on the western half of our map causing a saturated region to appear
cyan in our color image.

Our Perseus observations also have a significant amount of dust nebulosity. A
particularly bright region near the center of our map contains the IRAS source
03382+3145. To the southeast of this region is a cluster of bright red sources.
Based on these sources' red appearance in Figure \ref{fig:perseusrgb}, they
likely have an infrared excess suggesting they are young stellar objects. These
sources were not resolved by IRAS, but IRAS 03388+3139, a known infrared source
is located at this position. This cluster is identified by \citet{jorgensen06}.
Several other isolated infrared excess sources exist within the field. Just
to the east of our observed region is the well-known star-forming region IC348.

Finally, in Serpens we mapped a region surrounding ``Cluster B'', a group of
YSOs to the south of the Serpens core region and located approximately in the
middle of the c2d area. This region contains numerous YSOs. Two notable dark
patches show up in the mid-infrared \spitzer{} data, one to the southeast of
``Cluster B'' and the other to the northeast. Just to the south of our \jhks{}
maps is the Herbig Ae star VV Ser surrounded by nebulosity.

\section{ \label{sec:clouddatareduce} Data Reduction }

We reduced our \jhks{} data using PyRAF. Each FLAMINGOS field consisted of
multiple dithers with small offsets around a central position. To reduce our
data, we first subtracted a median filtered dark image from each dither.
Next, we flat-fielded the dithers by dividing out a median filtered dome
flat. The last, and most important, step was to subtract out the sky. This is
critical for IR observations where the sky is typically much brighter than the
astronomical objects of interest. We used a two-pass sky subtraction with a
median sky computed from the nearest observed dithers in time. We identified the
stars from the first pass, then masked them out in the second pass to produce an
improved median sky image.

The stars in each field were found using the IRAF task \texttt{daofind} and
\texttt{daophot} was used for the point spread function (PSF) photometry. We
visually inspected the results to remove false sources found by \texttt{daofind}
and also to add real sources missed by it. Finally, we used the 2MASS catalog to
correct the coordinates in each field and also to calibrate the photometry.

The \spitzer{} data were processed with the c2d data pipeline and are available
on the Spitzer Science Center's 
website\footnote{\url{http://ssc.spitzer.caltech.edu/legacy/c2dhistory.html}}.
Details on data reduction may be found in the c2d data delivery documentation,
also available on the website, and in several previously published papers 
\citep{padgett08,jorgensen06,rebull07,harvey06,harvey07a}.

\subsection{ \label{sec:reliability} Data Quality }

The mean difference in position between sources in our \jhks{} catalogs and in
the 2MASS catalogs is $0\farcs15$ and 95\% of our sources have a difference in
position of $\leq0\farcs6$ (around 2 pixels on the FLAMINGOS CCD). To test our
photometric error, we computed the difference in flux between sources present in
our Kitt Peak observations and the 2MASS catalogs. The resultant distribution is
Gaussian with $\sigma = 4-5$\%.

The c2d delivery documentation discusses three sources of uncertainty:
statistical, systematic, and absolute.  The first two of these are incorporated
into the photometric uncertainties listed in the c2d catalog.  These errors are
derived from the repeatability of flux measurements using the c2d pipeline. 
Since we are using the c2d data in this paper, we will adopt their
uncertainties. The c2d documentation lists a $4.6\%$ systematic error for IRAC
and $9.2\%$ for MIPS.  Combining the systematic and statistical errors, our
final photometric uncertainties are approximately $5\%$ in IRAC and $10\%$ in
MIPS.  The absolute uncertainties in the flux calibration are 1.5\% and 4\%,
respectively, for the IRAC and MIPS1 ($24\:\mu$m) bands. We obtained these
values from the Infrared Array Camera (IRAC) Data Handbook, Version 3.0 and the
Multiband Imaging Photometer for Spitzer (MIPS) Data Handbook, version 3.3.0. 
The absolute uncertainties are added in quadrature with the photometric errors
during source classification, but are not listed in the final catalogs since
they are smaller than the photometric errors.

Our 5 and $10\sigma$ detection limits are listed in Table \ref{tab:sigmalimits}.
For the \jhks{} bands our $10\sigma$ limits are 19.5, 18.8, and 17.7 magnitudes,
respectively. These limits are $\sim3.5$ magnitudes deeper than 2MASS in all
three bands.

\subsection{\label{sec:class} Extinction and Star Classification}

The standard c2d pipeline uses all available wavelengths to compute the 
extinction towards each star:

\begin{equation}
\label{eq:fit}
\log(F_{obs}(\lambda)/F_{model}(\lambda)) =
\log(k)-0.4\times C_{ext}(\lambda) \times A_V
\end{equation}

\noindent where $F_{model}(\lambda)$ is the stellar photosphere model, $k$ is
the scaling factor of the model for a particular star, and
\mbox{$C_{ext}(\lambda) \equiv A_{\lambda}/A_V$} is the ratio of extinction at
wavelength $\lambda$ to visual extinction from the dust extinction law. $k$ and
$A_V$ are derived from the linear fit of this equation by adopting stellar
photosphere and dust extinction models. The stellar photosphere models for
$K_s$--MIPS1 bands are based on the Kurucz-Lejeune models and come from the
SSC's ``Star-Pet''
tool\footnote{\url{http://ssc.spitzer.caltech.edu/tools/starpet}}. For the 2MASS
bands, c2d translated the observed $J-H$ and $H-K$ colors of stars
\citep{koornneef83} to fluxes relative to $K$ band and ignored the difference
between the $K$ and $K_s$ bands.  Sources that were a good fit to  this equation
were classified as reddened stars; those that did not fit were compared with
other templates to classify the sources.

In this paper, we slightly modified this procedure. First, the extinction, \ak,
was computed for each source using the \jhks{} bands and the NICER technique
\citep{lombardi01}. We also computed extinctions in \ak{} rather than the more
traditional $A_V$ so we can directly compare our results with those of other
authors. The NICER technique relies on the assumption of intrinsic values for
the $J-H$ and $H-K_s$ colors of stars. For this paper we adopted the intrinsic
colors $J-H = 0.63\pm0.16$ and $H-K_s = 0.21\pm0.14$. These values were based on
the average color for stars from off-cloud fields. Then, using Equation
\ref{eq:fit}, we identified the stars with the extinction held fixed.

We made this change because our primary goal in this paper is to compute the
extinction law at the IRAC and MIPS wavelengths. The method we use for deriving
the mid-infrared extinction law (\alak) is dependent on \ak{} so we cannot use
these wavelengths when computing \ak. However, we still need to choose an
appropriate extinction law. This will be used to compute \ak{} from the \jhks{}
bands and also for the source classification. For purposes of source
classification, we considered two dust models, the \citet{weingartner01} $R_V =
3.1$ and $R_V = 5.5$ models. The $R_V = 3.1$ model (hereafter WD3.1) is designed
to reproduce the extinction law of the diffuse interstellar medium; the $R_V =
5.5$ model (hereafter WD5.5) fits the observed law of denser regions. As we
discuss in the next section, we adopted the WD5.5 dust model for computing the
extinctions and source classifications.  The parameter $R_V = A_V/ (A_B - A_V)$
was historically used to characterize changes in the extinction law.

The advantage of using only the near-infrared bands to compute \ak{} is that the
extinction law in the near-infrared \jhks{} bands is very similar for both the
WD3.1 and WD5.5 models. In \S\,\ref{sec:beta-clouds} we will discuss possible
biases in our results attributable to our assumption about the near-infrared
extinction law.

\subsection{\label{sec:highreliable} High-Reliability Star Catalogs}

\begin{figure}

\plotone{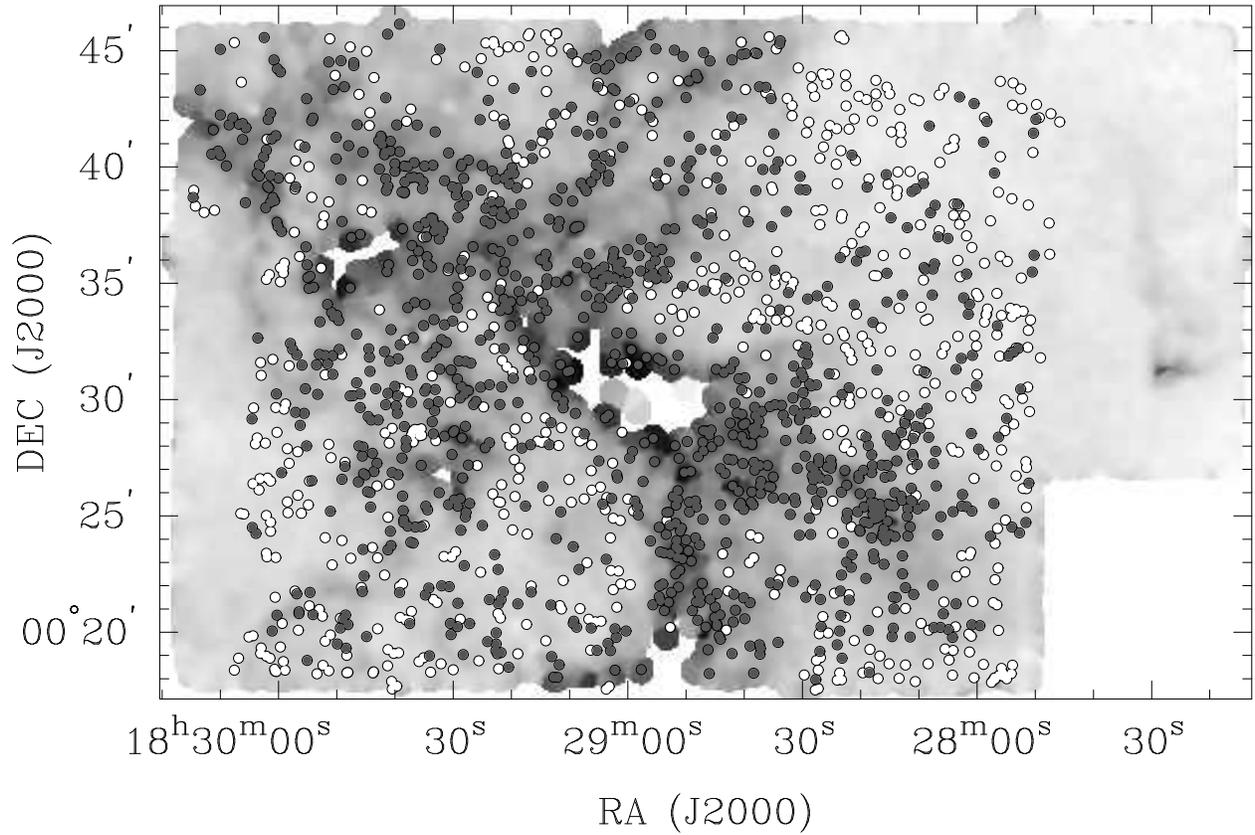}

\caption[Stars With Unusual Extinction Properties ]{\label{fig:onlyrv55}
Extinction map of Serpens (See \S\,\ref{sec:cloudext}) with sources identified
as stars using only the WD3.1 (white) or WD5.5 (dark gray) dust models. }

\end{figure}

We initially created high-reliability star catalogs by selecting only those
sources classified as stars with both the WD3.1 and WD5.5 dust models. This
selection excludes questionable stars that can affect our results, but may also
miss bona-fide stars that trace a true change in the extinction law. To check
whether this happens, we plotted in Figure \ref{fig:onlyrv55} those sources from
Serpens classified as stars using either WD3.1 or WD5.5, but not both. We found
that the sources classified as stars with WD5.5 are strongly associated with the
higher extinction regions while the WD3.1 stars are scattered randomly
throughout the observed regions. Furthermore, up to a third of the detections
that are only classified as stars using WD3.1 have a quality flag identifying
them as confused with a nearby source suggesting that many of these sources have
suspect photometry. In the remainder of the catalogs, $<1\%$ of the sources are
similarly confused.

For these reasons, we selected all sources identified as stars with the
WD5.5 dust model and computed line-of-sight extinctions based on this model. We
did not want a prominent absorption or emission line to bias our results.
Therefore, we excluded any sources classified as stars only when one of the
wavebands was dropped. Furthermore, because accurate extinction measurements are
essential to this paper, we required sources to have a detection $\ge 7
\sigma$ in each of the $JHK_s$ bands. Lastly, we `cleaned' our star catalogs to
remove suspected faint background galaxies that were mis-identified as stars.
This procedure will be discussed in detail in \S \ref{sec:faintgalaxy}. In
our final catalogs we have the following numbers of sources before (after)
cleaning: 2,403 (2,365) in Ophiuchus, 11,479 (11,280) in Perseus, and 49,712 
(49,485) in Serpens.

\subsection{ Young Stellar Objects and Background Galaxies }

In addition to stars, the c2d classification procedure also identifies Young
Stellar Object candidates (YSOc) and background galaxy candidates (Galc). Within
the regions we observed with deep \jhks{} and \spitzer{} observations, we
identified the following numbers of YSOc (Galc) sources: 25 (6) in Ophiuchus, 40
(25) in Perseus, and 74 (21) in Serpens. Of course, many more YSOc and Galc
objects are identified in the full cloud catalogs. Readers interested in the
YSO/Galc content of the full clouds should consult \citet{padgett08}
(Ophiuchus), \citet{jorgensen06} and \citet{rebull07} (Perseus), and
\citet{harvey07} (Serpens). All but fourteen of our YSOc sources have the same
identification in the c2d catalogs. Thirteen of the fourteen are attributable to
our modified source classification procedure. Our method tends to be
conservative when classifying stars, thus ambiguous sources may become YSOs. In
contrast, the c2d pipeline is more conservative when selecting YSOc sources, so
ambiguous cases may become stars. The fourteenth source is classified as a YSOc
in our catalogs, but as a Galc by c2d. This is because it is detected in our
deep \jhks{} data but not by the 2MASS survey used by c2d. With our data, the
c2d pipeline also classifies this source as a YSOc. In the c2d catalogs, this is
source SSTc2d J034214.9+314758. All of our Galc sources are also identified as
such by c2d.

\subsection{ \label{sec:faintgalaxy} Misidentified Background Galaxies }

\begin{figure}

\plotone{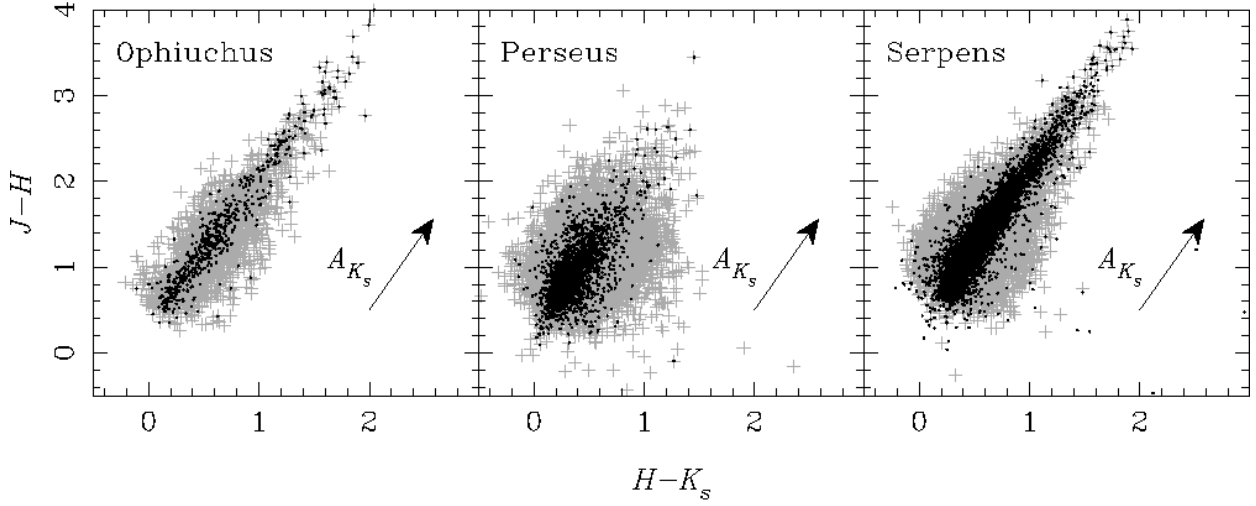}

\caption[$J-H$ vs.\ $H-K_s$ for the Clouds ]{ \label{fig:ahak-bad} Near-infrared
color-color diagrams for Ophiuchus, Perseus, and Serpens. Gray crosses have
$K_s$ magnitude $> 15$ while the black points are brighter. The extinction
vector shown is for the WD5.5 model.}

\end{figure}

Figure \ref{fig:ahak-bad} is a color-color diagram showing $J-H$ versus $H-K_s$
for our three clouds. Those sources brighter than 15th magnitude at $K_s$ are
shown in black while fainter sources are gray crosses. We expect to see
extincted stars following a reddening vector that stretches the data points into
a line with some scatter, as shown in the figure. We do see this trend with the
black points in all three clouds and also with the gray crosses in Ophiuchus and
Serpens. However, this trend is not clear in Perseus where many gray sources
with $H-K_s \gtrsim 0.7$ and $J-H \lesssim 1.4$ do not appear to follow the
reddening vector.

What is the nature of these sources? One possibility is that these are merely
faint stars with bad photometry. It is true that these sources are fainter than
15th magnitude at $K_s$, but we are already using our high reliability catalogs
to ensure good photometry. Upon further examination, we found many of these
anomalous sources are only detected in the $J$-IRAC2 bands. Another possibility,
therefore, is that these sources are not really stars, even though they have
been classified as such. Perseus is located at Galactic longitude $l =
160^\circ$, almost in the Galactic anti-center direction, while Ophiuchus and
Serpens are more towards the Galactic center at $l = 353^\circ$ and $l =
30^\circ$, respectively. With these Galactic longitudes and given that these
anomalous sources are seen only in Perseus, we suspect these sources are likely
misidentified background galaxies. Our hypothesis is supported by the recent
work of \citet{foster08}. They identified galaxies from near infrared \jhks{}
data and found that their galaxies were located in the same region of the $J-H$
vs.\ $H-K_s$ color-color diagram as our suspected galaxies.

Our goal in this section is to develop a set of criteria that we will use to
separate these suspected faint background galaxies from our actual stars. It is
likely that any procedure will remove some bona fide stars and embedded objects
in addition to background galaxies. We want our criteria to be stringent enough
so that the final sample of sources selected for removal will be strongly
enriched in background galaxies but also contain a minimum number of actual
stars. We will then use this procedure to create purified star catalogs.

\begin{figure}

\plotone{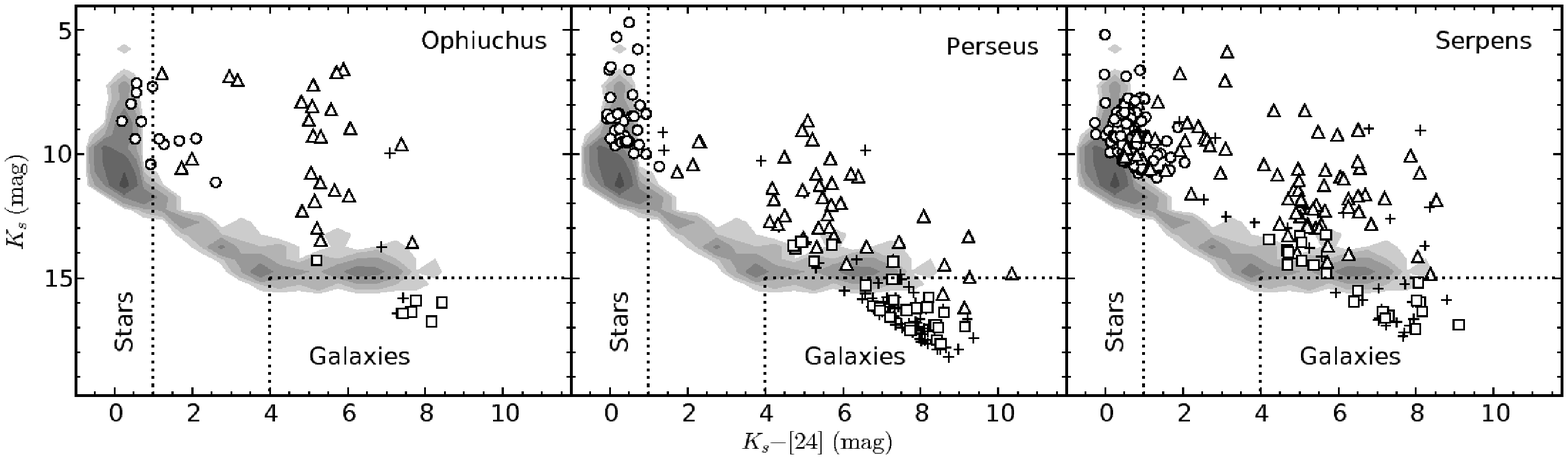}

\caption[ Color-Magnitude diagram of Clouds ]{\label{fig:cloudk24} $K_s$ vs.
$K_s - [24]$ plot for our three clouds. We used four different symbols to
correspond to different source classifications: stars are circles, background
galaxies are squares, `YSO candidates' are triangles, and plus signs denote all
other classifications. The shaded contours are from the c2d processed catalog of
part of ELAIS N1 \citep{surace04}. The dashed lines denote the cutoffs we
used for selecting ``known'' stars and background galaxies.}

\end{figure}

To determine whether the sources with anomalous colors really are background
galaxies, we need a group of ``known'' stars and background galaxies with which
we can compare. To make these ``known'' samples, we plotted $K_s$ vs. $K_s -
[24]$ for our three cloud regions in Figure \ref{fig:cloudk24}. We only plotted
those sources with S/N$ \ge 3\sigma$ at $24\:\mu$m in addition to the earlier
cut ($7\sigma$) we made for uncertainties in $JHK_s$ photometry. The shaded
contours in each panel are the SWIRE data of ELAIS N1 \citep{surace04} as
processed by the c2d team. The ELAIS N1 region is near the north Galactic pole,
and therefore the data should contain nothing but stars and background galaxies
making them useful for selecting these types of objects. Based on Figure
\ref{fig:cloudk24} we selected two groups of sources: those with $K_s - [24] \le
1$ and those sources with $K_s \ge 15$ and $K_s - [24] \ge 4$. The first group,
with no color excess, contains stars, while the second population consists of
background galaxies. We will use these two groups as our ``known'' samples.

\begin{figure}

\plotone{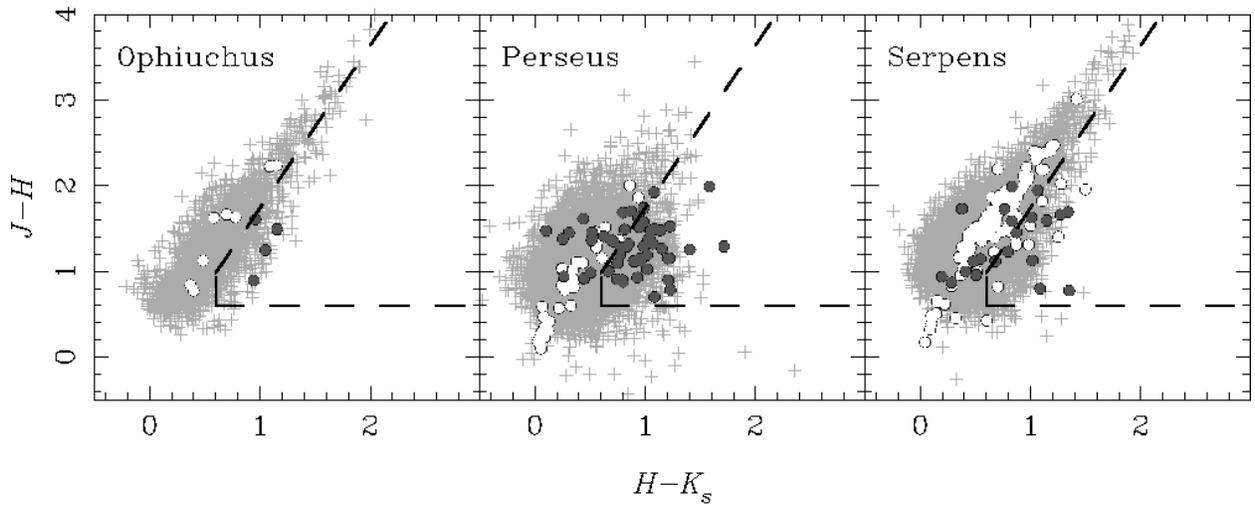}

\caption[$J-H$ vs.\ $H-K_s$ With Stars and Galaxies ]{ \label{fig:ahak-withgal}
Near-infrared color-color diagrams for Ophiuchus, Perseus, and Serpens. The
white and dark gray circles are the `known' stars and background galaxies,
respectively. The dashed lines select those sources with $J-H \ge 0.6$, $H-K_s
\ge 0.6$, and $J-H \le 1.9 \times (H-K_s) - 0.16$. }

\end{figure}

In Figure \ref{fig:ahak-withgal} we overlaid the ``known'' stars (white circles)
and background galaxies (dark gray circles) on our $J-H$ vs. $H-K_s$ color-color
diagram. These two classes of sources are roughly separated. The stars chiefly
lie along the stellar reddening vector, while the background galaxies appear
below and to the right of this vector, exactly where our anomalous sources appear
in Perseus. As our first selection criterion, we drew dashed lines to separate
the stars from the possible galaxies. Note that the separation between the two
is not perfect, but does establish that sources with $J-H \ge 0.6$, $H-K_s \ge
0.6$ and to the right of the sloped line are more likely to be galaxies rather
than stars. The equation for the sloped line is: $J-H = 1.9 \times (H-K_s) -
0.16$.

\begin{figure}

\plotone{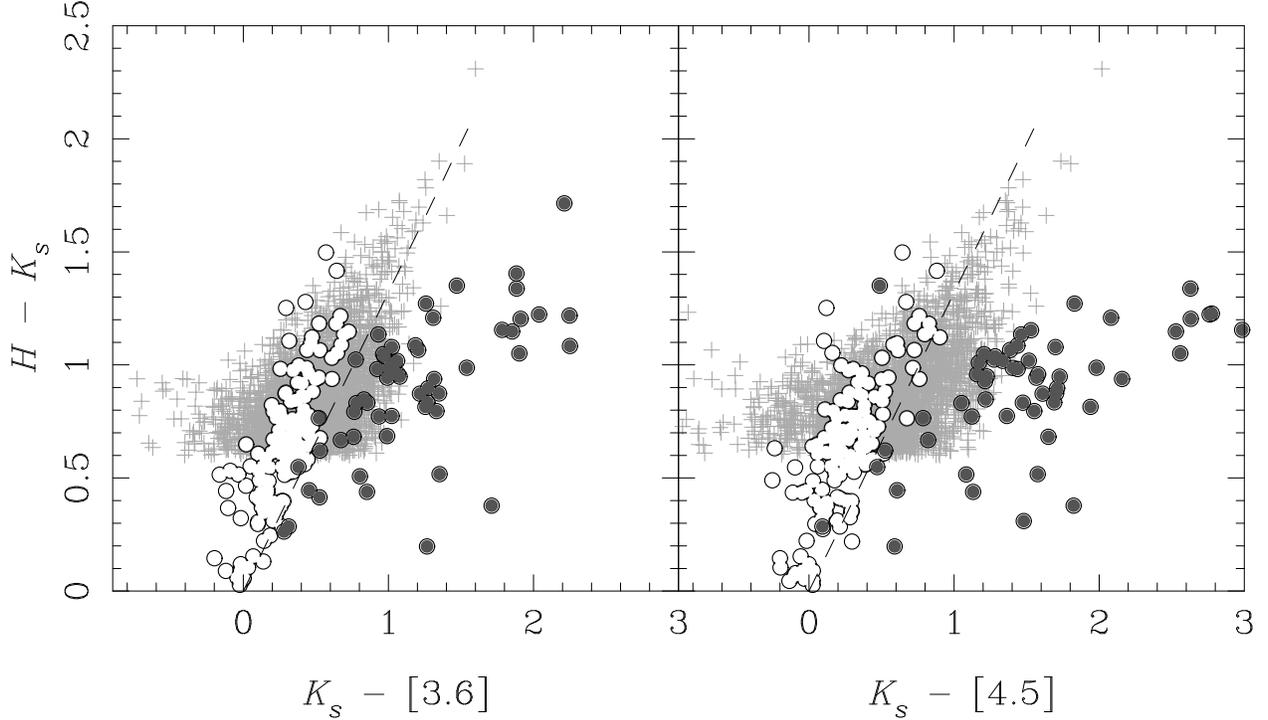}

\caption[ Color-Color Diagram Showing Background Galaxies ]{ \label{fig:cc-bad}
$H-K_s$ vs.\ $K_s - [3.6]$ (left) and vs.\ $K_s - [4.5]$ (right). The gray
crosses are the sources selected by the dashed lines in Figure
\ref{fig:ahak-withgal} for Ophiuchus, Perseus, and Serpens. The white and dark
gray circles are the `known' stars and background galaxies, respectively. The
dashed line in each panel has the equation $H-K_s = 1.32x$ where $x$ is either
$K_s - [3.6]$ or $K_s - [4.5]$. }

\end{figure}

As noted earlier, approximately three-fourths of the objects in our clouds that
lie in the region selected by our first criterion are faint enough that they are
only detected in the $J$-IRAC2 wavebands. Therefore, in Figure \ref{fig:cc-bad}
we plotted $H-K_s$ vs.\ $K_s - [3.6]$ and vs.\ $K_s - [4.5]$. The gray crosses
are those sources selected in all three clouds from Figure
\ref{fig:ahak-withgal} by the dashed lines. Again, white circles are stars and dark gray circles are
background galaxies. In both of these plots, the stars and galaxies are clearly
separated. We defined a single line to use for both plots: $H-K_s = 1.32x$ where
$x$ is either $K_s - [3.6]$ or $K_s - [4.5]$.

We combined these two selections to reduce the number of background galaxies in
our data. Starting from the high-reliability ``star'' catalogs, we removed all
sources satisfying the following two criteria:

\begin{enumerate}

\item $J-H \ge 0.6$, $H-K_s \ge 0.6$, and $J-H \le 1.9 \times (H-K_s) - 0.16$

\item $H-K_s \le 1.32 \times (K_s - [3.6])$ and $H-K_s \le 1.32 \times 
(K_s - [4.5])$

\end{enumerate}

Our cleaning procedure removed 1.6\% of the sources in Ophiuchus, 1.8\% in
Perseus, and 0.5\% in Serpens.

\section{ \label{sec:cloudext} Extinction }

As mentioned in \S\,\ref{sec:class}, we computed line-of-sight extinctions,
\ak, to each star using the NICER technique \citep{lombardi01}. These
line-of-sight (LOS) \ak{} values give us a randomly distributed sampling of the
true \ak{} in each cloud. To convert these LOS extinctions into a uniform map,
we overlaid a grid on top of the data to represent the cells in the final
extinction map. At each grid position, we used all of the LOS extinctions
within a given radius. The extinction value in that cell is then the average of
the individual extinctions, weighted both by uncertainty and distance from the
center of the cell using a Gaussian weighting function.

For each cell we chose an integration radius equal to the full width at half
maximum, or $2.3548\sigma$. All that remains then is to choose the appropriate
resolution for our maps. To obtain good statistical accuracy for each cell, we
used a resolution that yielded on average $20+$ stars in each cell. For
Ophiuchus and Perseus, this corresponds to $90\arcsec$ resolution while in
Serpens, which is towards the Galactic center, we were able to make a
$30\arcsec$ resolution map and still have on average 36 stars per cell.

\begin{figure}

\plotone{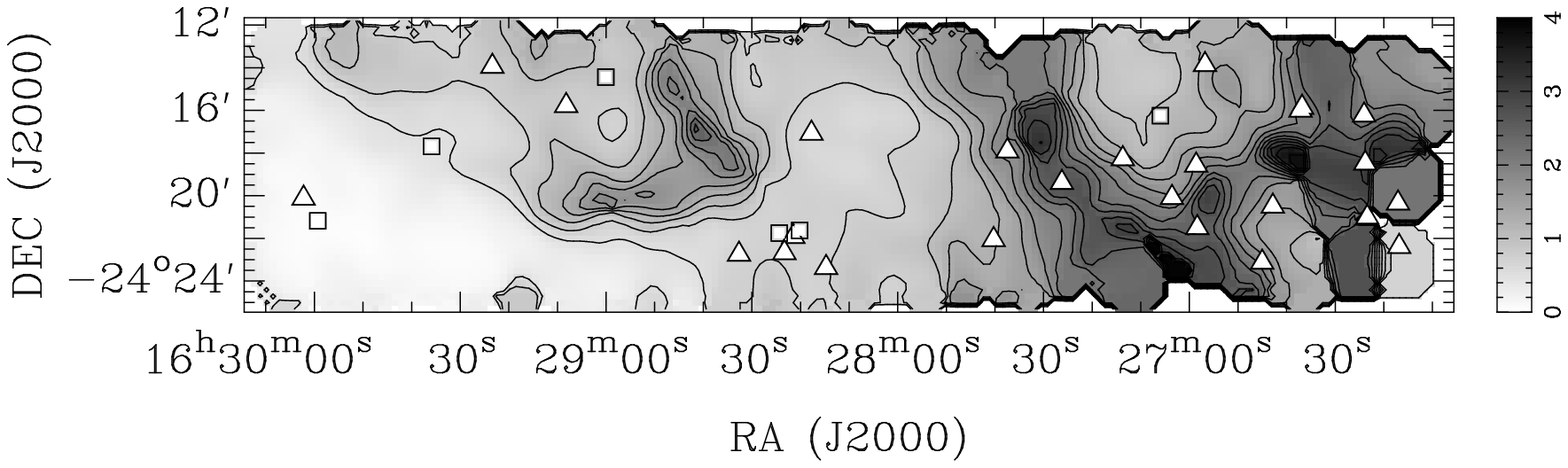}

\caption[ Extinction map of Ophiuchus ]{\label{fig:ophak} \ak{} map of our
observed region within the Ophiuchus cloud. The map has a resolution of
$90\arcsec$. The contours start at $A_{K_s} = 0.5$ in steps of 0.25 ($5\sigma$).
We plotted all the YSOc and Galc objects as triangles and squares,
respectively. The maximum \ak{} value in the map is 4.0 magnitudes. }

\end{figure}

\begin{figure}
\epsscale{0.9}
\plotone{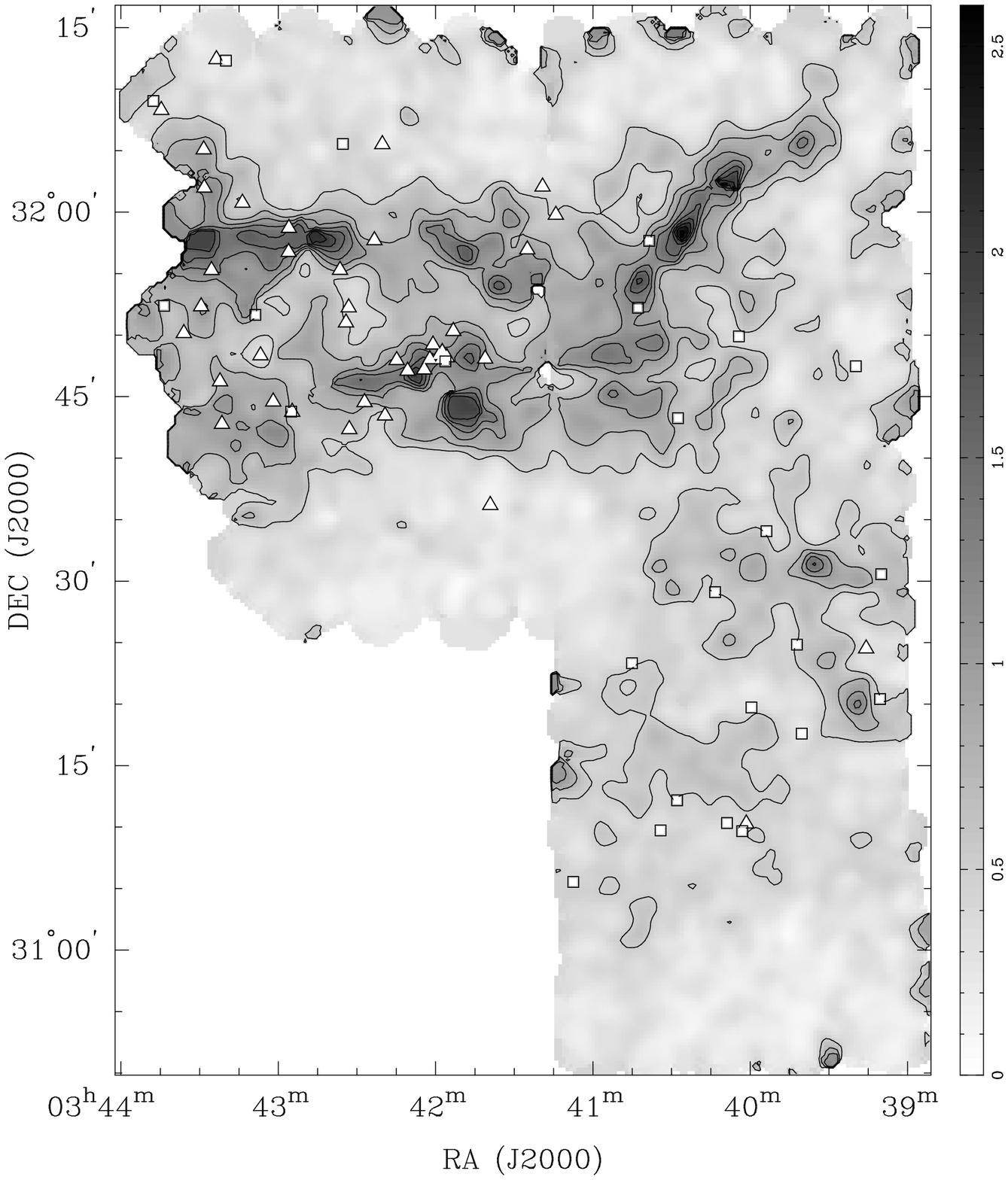}

\caption[ Extinction map of Perseus ]{\label{fig:perseusak} \ak{} map of our
observed region within the Perseus cloud. The map has a resolution of
$90\arcsec$. The contours start at $A_{K_s} = 0.5$ in steps of 0.25 ($5\sigma$).
We plotted all the YSOc and Galc objects as triangles and squares,
respectively. The maximum \ak{} value in the map is 2.6 magnitudes. }

\end{figure}

\begin{figure}
\epsscale{1}
\plotone{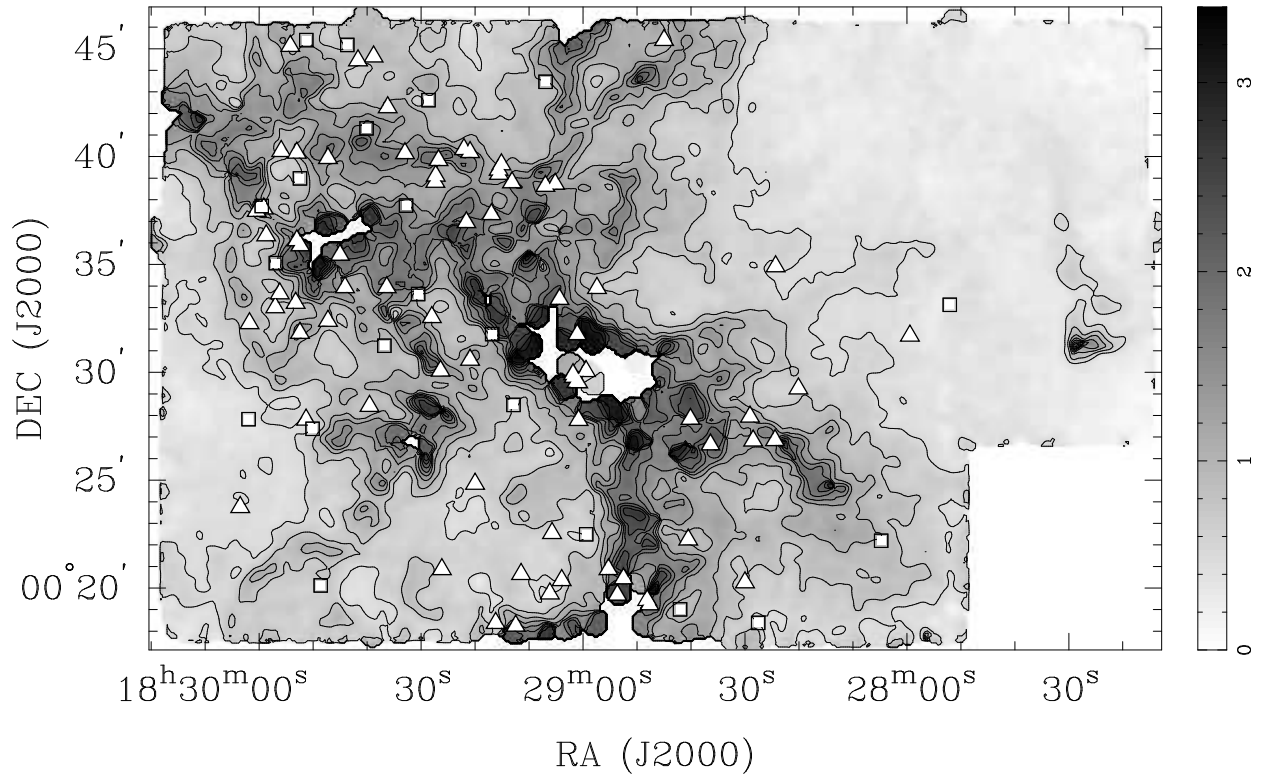}

\caption[ Extinction map of Serpens ]{\label{fig:serpensak} \ak{} map of our
observed region within the Serpens cloud. The ``holes'' in the map correspond to
regions with no data eitherbecause they are too dense or because the only
sources present were not classified as stars. The map has a resolution of
$30\arcsec$. The contours start at $A_{K_s} = 0.5$ in steps of 0.25 ($5\sigma$).
We plotted all the YSOc and Galc objects for Serpens as triangles and
squares, respectively. The maximum \ak{} value in the map is 3.4 magnitudes. }

\end{figure}

Foreground stars, which are correctly assigned a small \ak, bias the average
extinction for a given cell. We dealt with foreground stars as follows: for
each cell with more than 2 stars, we computed the mean and median \ak{} value.
If these two statistics differed by more than 25\%, then we dropped the source
with the lowest \ak{} and recomputed the mean and median. However, if the
difference between the old and new mean extinction is $<25\%$, then we
re-added the dropped source to our catalog. After we identified all the  sources
to drop, we recomputed the extinctions in every cell.  Our extinction maps of
Ophiuchus, Perseus, and Serpens are shown in Figures \ref{fig:ophak} -
\ref{fig:serpensak}.

Our Ophiuchus region has the largest dynamic range of \ak{} in our clouds, up to
4 magnitudes. A loop of extinction is located at $\sim16^h28^m30^s$ in
addition to the high extinction region L1688 on the western side of the map.
Two-thirds of the YSOc sources are located in this half.

Our Perseus extinction map contains dust filaments in the upper half of the
image and little extinction above the $A_{K_s} = 0.5$ level elsewhere in the
map. A cluster of YSOs identified as IRAS 03388+3139 is located at RA
$\sim3^h42^m$ DEC $\sim31^\circ 47\arcmin$. Furthermore, the high extinction
regions appear to be correlated with the darker patches in the color image. The
dust lane that crosses the nebulous region of IRAS 03382+3145 ($\sim3^h41^m30^s$
$+31^\circ55\arcmin$) shows up as a high extinction filament in our extinction
map. Almost all the Galc sources are in the low-extinction southern half of the
map.

The Serpens extinction map has a higher resolution, $30\arcsec$, than either the
Ophiuchus or Perseus maps. Two regions show up as holes in our maps because of a
lack of background stars. The larger hole in the center of the map contains the
star-forming region known as Cluster B while the extinction hole west of Cluster
B shows up as a dark patch even at $24\:\mu$m.

\section{ \label{sec:dustprop-clouds} Dust Properties }

The WD3.1 and WD5.5 models are nearly identical in the near-infrared \jhks{}
bands. We will exploit this constancy of the extinction law in the near-infrared
by using the extinctions computed from the \jhks{} bands and then extrapolating
the extinction law to the \spitzer{} wavebands.

Starting from our basic equation relating flux and extinction,

\begin{equation}
\log(F_{obs}(\lambda)/F_{model}(\lambda)) =
\log(k)-0.4\times C_{ext}(\lambda) \times A_{K_s}
\end{equation}

\noindent we can re-arrange this to solve for $C_{ext}$, defined as
$A_{\lambda}/A_{K_s}$:

\begin{equation}
C_{ext}(\lambda) = \frac{2.5}{A_{K_s}}[\log(k) -
\log(F_{obs}(\lambda)/F_{model}(\lambda))]
\end{equation}

Since we are only interested in \emph{differences} in the extinction law, we
subtracted $C_{ext}(K_s)$ to eliminate the need for $k$, the scaling factor. 
Furthermore, because $C_{ext}(K_s) \equiv 1$ and $F_{model}(K_s) \equiv 1$ (all
stellar models are scaled relative to $K_s$), the equation simplifies to:

\begin{equation}
\label{eq:cext}
C_{ext}(\lambda) = \frac{2.5}{A_{K_s}}\left[ \log(F_{obs}(K_s)/F_{obs}(\lambda))
   + \log F_{model}(\lambda)\right] + 1
\end{equation}

For $F_{model}(\lambda)$ we use the average stellar model derived in Appendix
\ref{sec:avgmodel}.

\subsection{\label{sec:chi2clouds} \chisq{} in the Clouds }

Our goal in this section is to construct two-dimensional maps of the changes
in the dust properties across each cloud. We will do this by computing a reduced \chisq{} value along
each line-of-sight and then applying the same method we used to create \ak{}
maps, except this time our ``extinction'' values will be \chisq{} values. We
define \chisq{} as the sum of the difference between our computed extinction law
and a theoretical model over the \spitzer{} IRAC wavebands ($3.6 - 8\:\mu$m).
The reduced \chisq{} is then simply \chisq{} divided by the number of bands
summed minus 1, since we have 1 free parameter in the model, namely \ak:

\begin{equation}
\label{eq:chi2}
\chi^2 = \frac{1}{n-1}\sum_\lambda^n \left( \frac{C_{ext}^{obs}(\lambda) - 
C_{ext}^{model}(\lambda)}{\sigma_\lambda}\right)^2
\end{equation}

$C_{ext}^{obs}(\lambda)$ is computed from Equation \ref{eq:cext},
$C_{ext}^{model}$ is the extinction law for a given dust model, and
$\sigma_\lambda$ is the uncertainty in $C_{ext}^{obs}$. We used both the
WD3.1 and WD5.5 dust models in creating our \chisq{} maps. Note that we only
summed over the IRAC bands. As we will see later, the extinction law at
$24\:\mu$m is almost always significantly higher than either the WD3.1 or WD5.5
extinction models (\S\,\ref{sec:extlawmips}). We do not want to bias our results
based on those data so we excluded this data point when calculating the
reduced \chisq. Finally, we excluded any negative values for
$C_{ext}^{obs}(\lambda)$ as unphysical.

\begin{figure}

\plotone{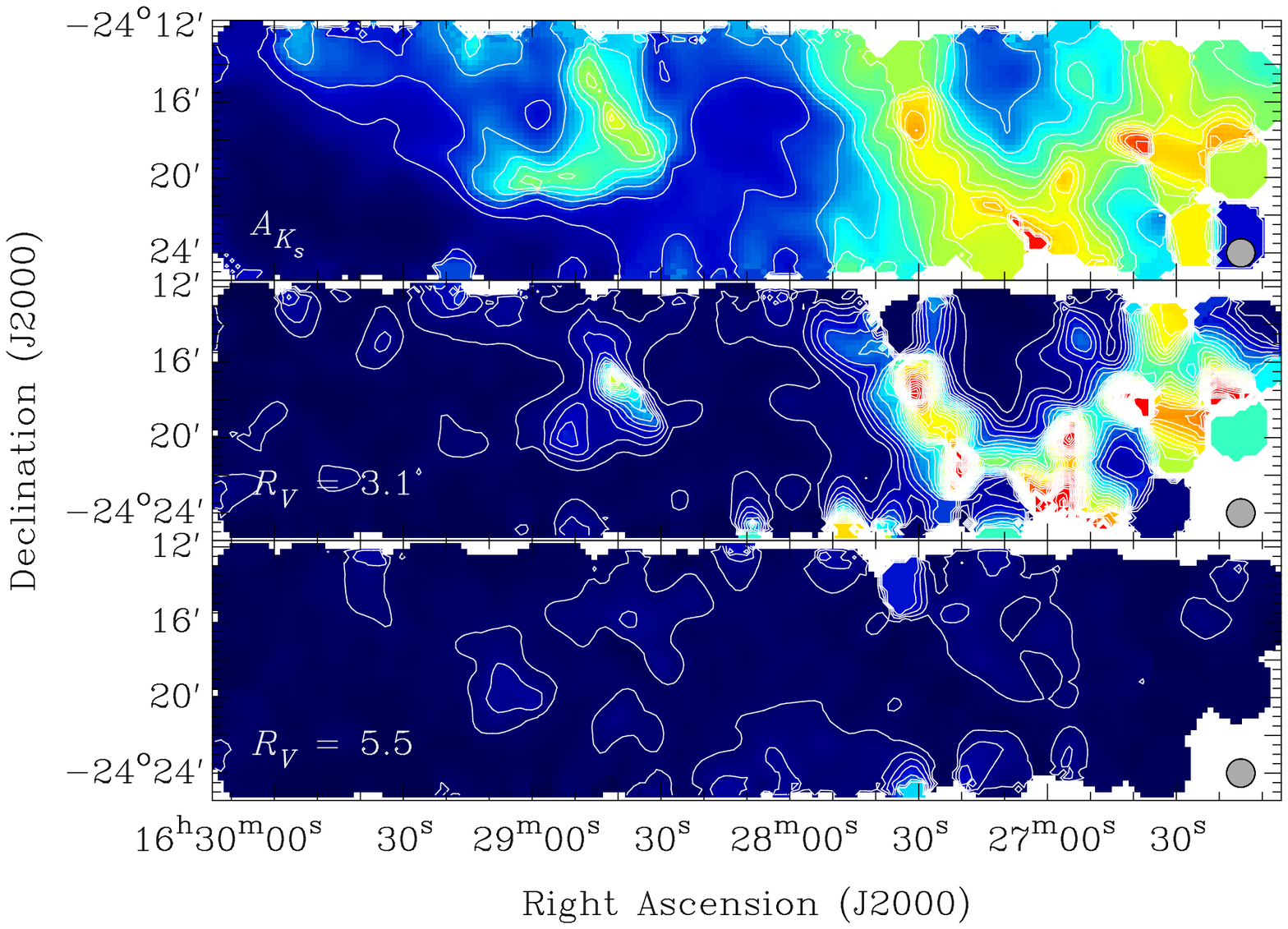}

\caption[ \chisq{} map of Ophiuchus ]{\label{fig:ophchi2} Map of the \ak{} and
\chisq{} values in Ophiuchus. The top panel shows the extinction map with
contours starting at 0.5 magnitudes in steps of 0.25 mag ($5\sigma$). The middle
and bottom panels show the \chisq{} maps for the same region made assuming
either the WD3.1 (middle) or WD5.5 (bottom) extinction laws.  Contours start at
$\chi^2 = 4$ in steps of 4. }

\end{figure}

\begin{figure}

\plotone{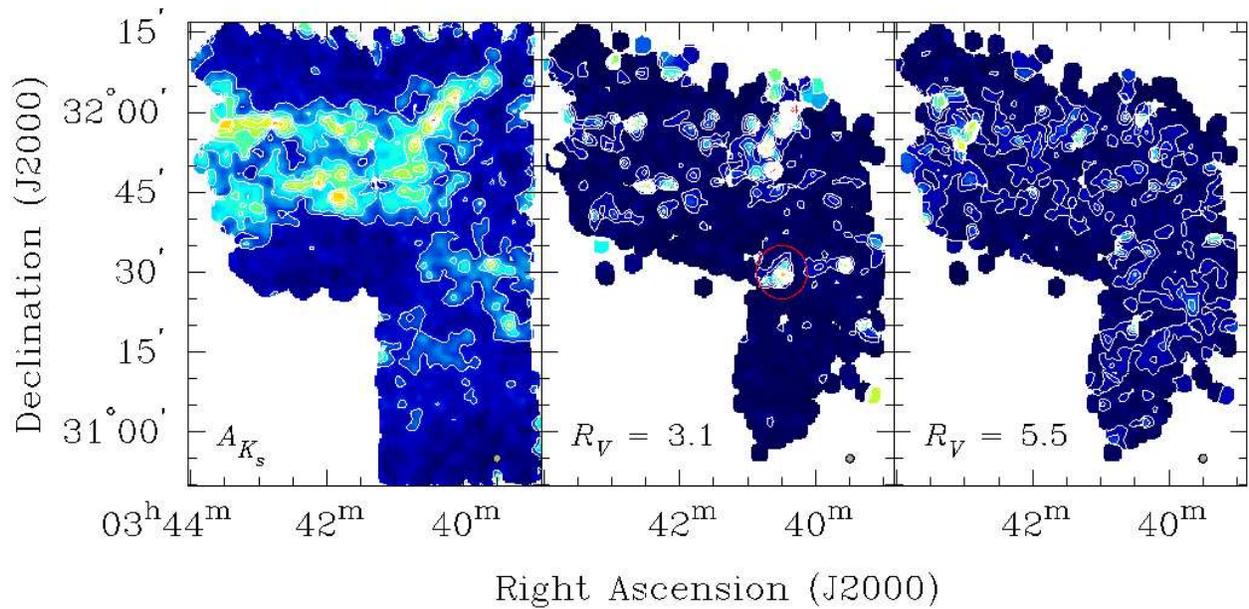}

\caption[ \chisq{} map of Perseus ]{\label{fig:perseuschi2} Map of the \ak{} and
\chisq{} values in Perseus. The left panel shows the extinction map with
contours starting at 0.5 magnitudes in steps of 0.25 mag.\ ($5\sigma$). The
middle and right panels show the \chisq{} maps for the same region made assuming
either the WD3.1 (middle) or WD5.5 (right) extinction laws. Contours start at 
$\chi^2 = 4$ in steps of 4. The region circled in red is discussed in
\S\S\,\ref{sec:chi2clouds} and \ref{sec:perseuschi2}.}

\end{figure}

\begin{figure}
\epsscale{0.65}
\plotone{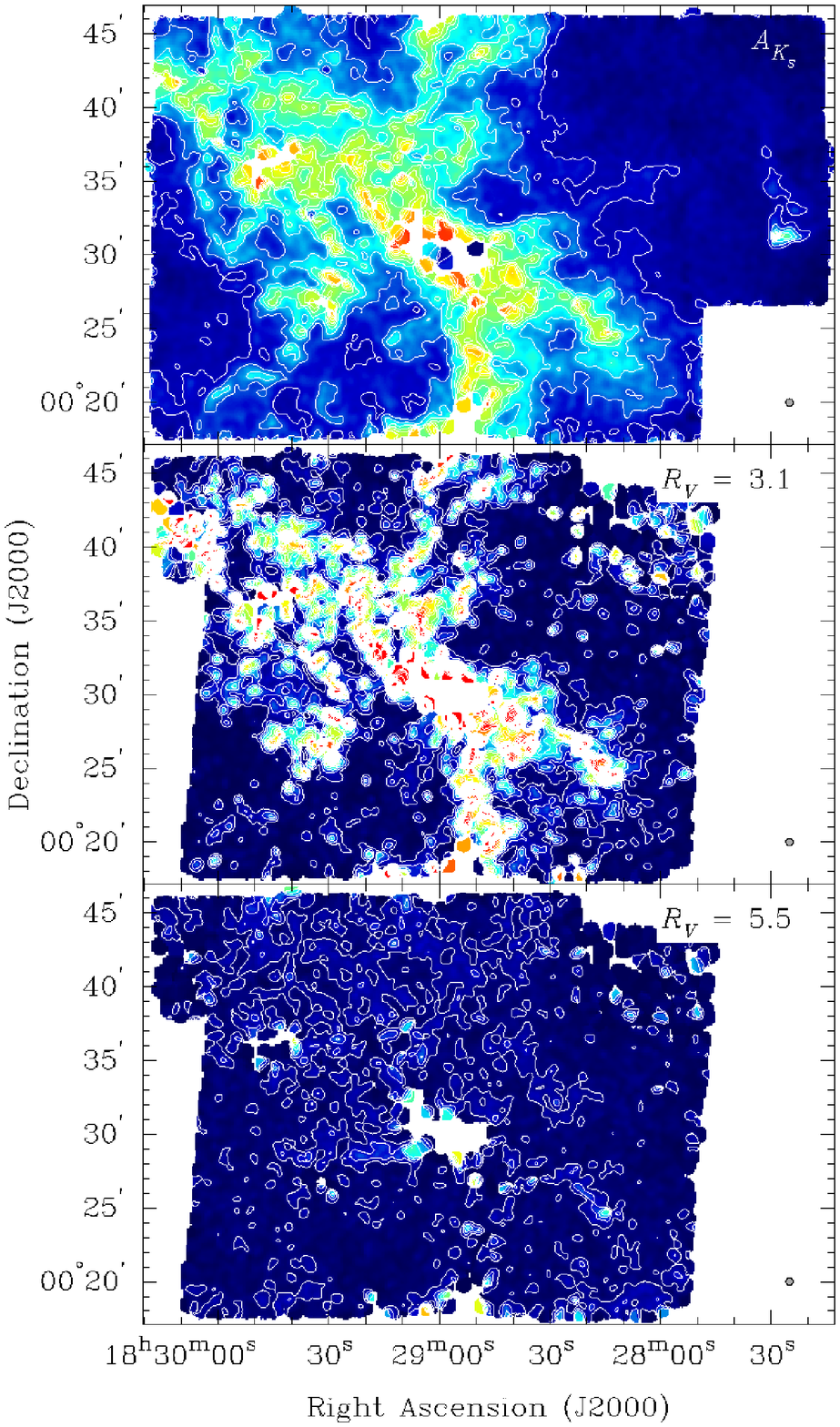}

\caption[ \chisq{} map of Serpens ]{\label{fig:serpenschi2} Map of the \ak{} and
\chisq{} values in Serpens. The top panel shows the extinction map with contours
starting at 0.5 magnitudes in steps of 0.25 mag. ($5\sigma$). The middle and
bottom panels are the \chisq{} maps for the same region made assuming either the
WD3.1 (middle) or WD5.5 (bottom) extinction laws. Contours start at $\chi^2 = 4$
in steps of 4. }

\end{figure}

\begin{figure}

\plotone{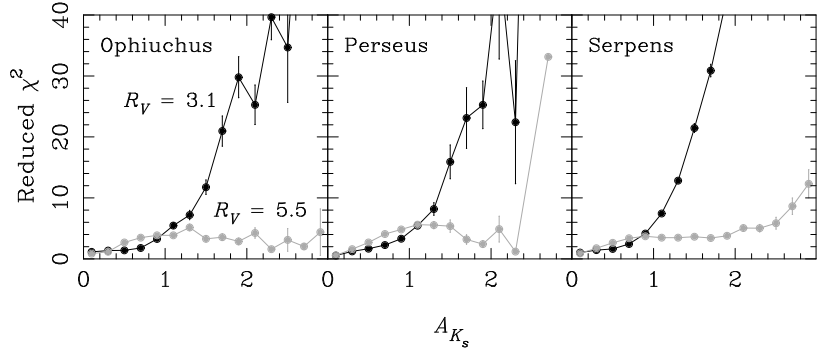}

\caption[ \chisq{} versus \ak{} For the Clouds ]{\label{fig:clouds-chi2ak}
\chisq{} versus \ak{} for the clouds we observed. The WD3.1 \chisq{} values are
black, while the WD5.5 \chisq{} values are light gray. The error bars shown are
the standard deviations of the mean \chisq{} value in each bin. }

\end{figure}

We created maps of the \chisq{} values using the same process we used for
extinction mapping and with the same resolution. Our \chisq{} maps for
Ophiuchus, Perseus, and Serpens are shown in Figures \ref{fig:ophchi2} -
\ref{fig:serpenschi2}. We drew our contours starting at $\chi^2 = 4$ because we
observe a definite transition between the WD3.1 and WD5.5 dust models at this
approximate \chisq{} value (see Figure \ref{fig:clouds-chi2ak}).  Statistically,
$\chi^2 = 4$ would arise by chance about $5\%$ of the time for $n = 2$.

In broad terms, high extinction regions are correlated with high $R_V = 3.1$
\chisq{} values. This same correlation is not seen with $R_V = 5.5$.  To make a
quantitative comparison between \chisq{} and \ak{}, we binned our data in \ak{}
and for each bin determined the average \chisq{} with both the WD3.1 and WD5.5
models. Our results are shown in Figure \ref{fig:clouds-chi2ak}. The $R_V = 3.1$
\chisq{} curve is shown in black while $R_V = 5.5$ is shown in gray. At low
extinction, $A_{K_s} \lesssim 1$, the $R_V = 3.1$ and $5.5$ curves are very
similar suggesting it is difficult to distinguish between the dust models with
our method.  Above $A_{K_s} \approx 1$, the $R_V = 3.1$ curve rises sharply
while $R_V = 5.5$ does not.

Our increase in $R_V = 3.1$ \chisq{} values in denser regions means the WD3.1
model does not fit the data in these regions.  The WD5.5 model, which includes
larger dust grains, produces much smaller \chisq{} values, suggesting grain
growth is occurring in the denser regions.  One prominent region within
the Perseus $R_V = 3.1$ \chisq{} map does not fit into this picture.  This
region is located at $\sim3^\mathrm{h}40.5^\mathrm{m}$ $+31^\circ 30\arcmin$ and
circled in red in Figure \ref{fig:perseuschi2}. This region has a peak \chisq{}
value of 39, but since it is non-circular and larger than the integration
radius, it seems unlikely that a single bad data point is affecting the
\chisq. The puzzling thing about this region is that the extinction is
fairly unremarkable suggesting that grain growth may not be causing the increase
in \chisq. This region will be examined in more detail in
\S\,\ref{sec:perseuschi2}.

\subsection{ \label{sec:extlawclouds} The Mid-Infrared Extinction Law }

\begin{figure}

\plotone{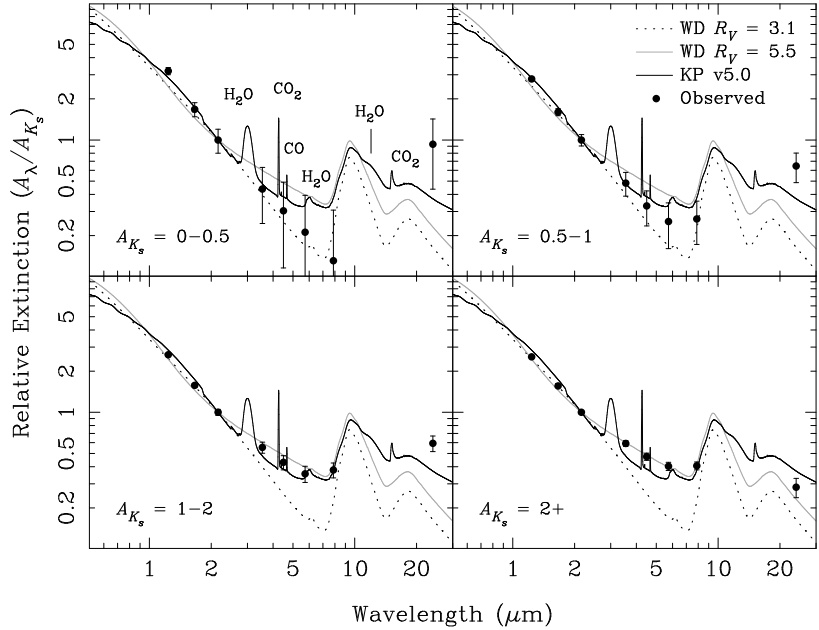}

\caption[ Extinction Law in Ophiuchus ]{ \label{fig:ophaklaw} The extinction law
in Ophiuchus in four different \ak{} ranges: 0-0.5, 0.5-1, 1-2, $>2$. The data
points are the weighted average observed extinction law computed from all
sources within the specified \ak{} range where the errorbars are the minimum
uncertainty due to systematic errors in measuring the fluxes. We plotted
three different dust models for comparison: \citet{weingartner01} $R_V = 3.1$
and 5.5 (dotted, gray lines), and KP v5.0 (Pontoppidian et al., in prep) in
black. Using \citet{gibb04} as a reference, we identified the ice features
in KP v5.0. }

\end{figure}

\begin{figure}

\plotone{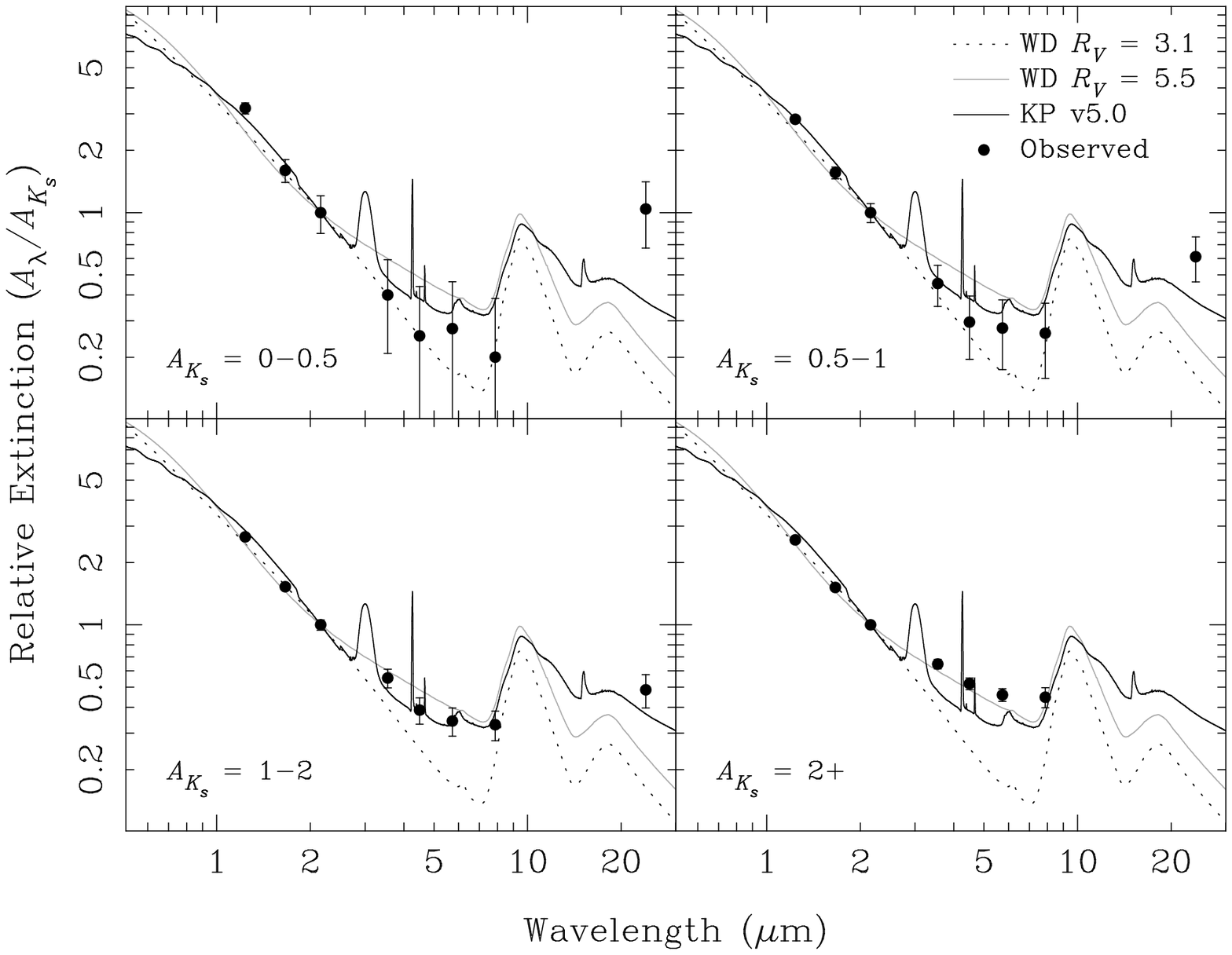}

\caption[ Extinction Law in Perseus ]{ \label{fig:perseusaklaw} The extinction
law in Perseus in four different \ak{} ranges: 0-0.5, 0.5-1, 1-2, $>2$.
See Figure \ref{fig:ophaklaw} for a description of the data points, curves, and
errorbars.}

\end{figure}

\begin{figure}

\plotone{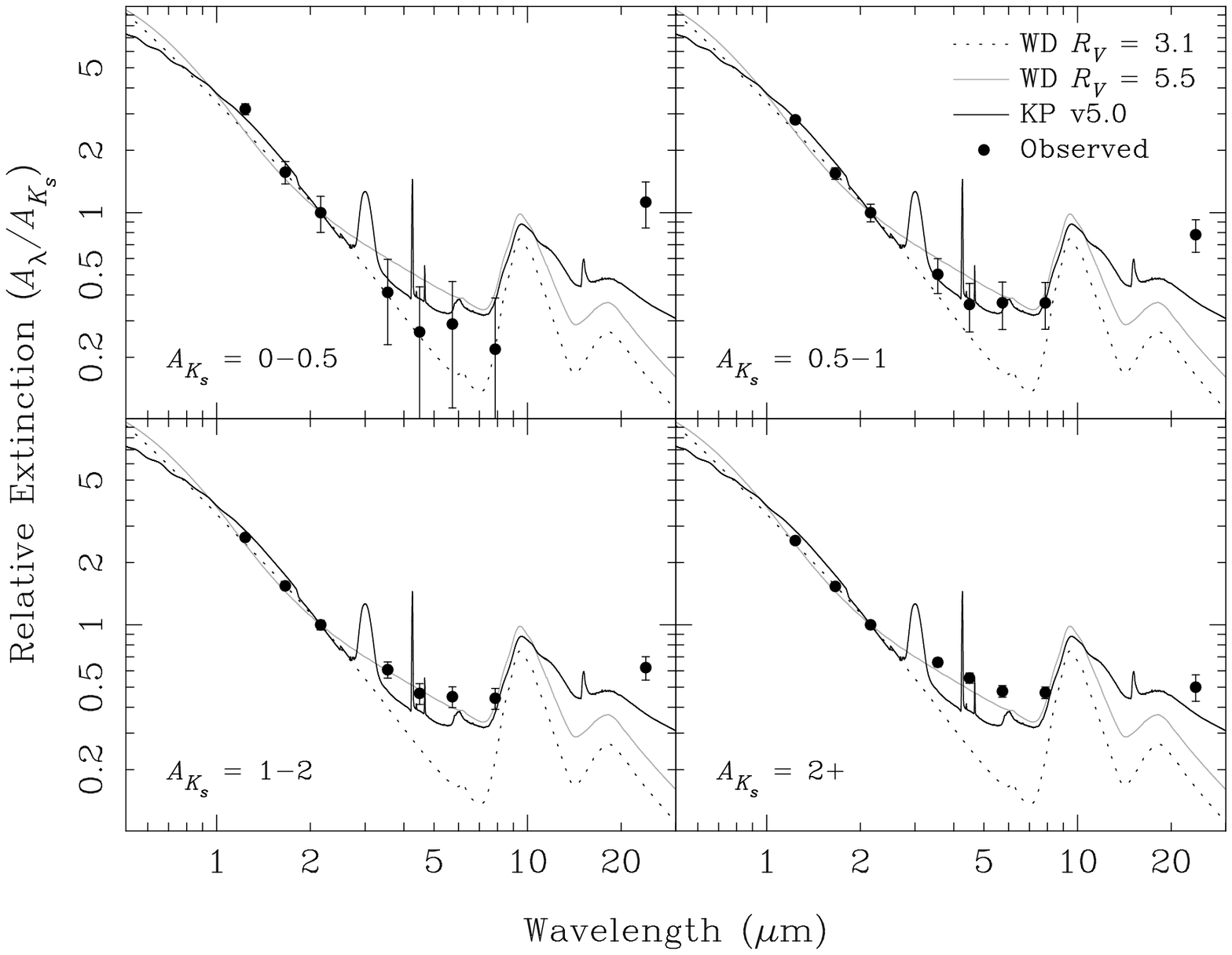}

\caption[ Extinction Law in Serpens ]{ \label{fig:serpensaklaw} The extinction
law in Serpens in four different \ak{} ranges: 0-0.5, 0.5-1, 1-2, $>2$.
See Figure \ref{fig:ophaklaw} for a description of the data points, curves, and
errorbars.}

\end{figure}

We have line-of-sight measurements of \ak{} and the extinction law for every
star within our clouds. In the previous section, we turned these measurements
into maps of the \chisq{} deviation from the WD3.1 or WD5.5 extinction laws. Now
we will examine the extinction law as a function of wavelength and \ak. We 
started by binning our line-of-sight measurements into four \ak{} ranges from
low to high: $0 < A_{K_s} \le 0.5$, $0.5 < A_{K_s} \le 1$, $1 < A_{K_s} \le 2$,
and $A_{K_s} \ge 2$. For each bin, we excluded any negative $C_{ext}^{obs}$
values as unphysical and computed the weighted average $C_{ext}(\lambda)$. If we
had used these negative $C_{ext}^{obs}$ values, then our extinction law for $0 <
A_{K_s} \le 0.5$ would be $A_{\lambda}/A_{K_s} \approx 0.02$ for IRAC2-4
($4.5-8\:\mu$m), suggesting the dust causes almost no extinction for these 
wavelengths and far below the predictions of dust models.  The impact of 
negative $C_{ext}^{obs}$ values on our other extinction bins is negligible. Our
results for our three clouds may be seen in Figures \ref{fig:ophaklaw} -
\ref{fig:serpensaklaw}. The errorbars in each bin represent the minimum
uncertainty due to systematic errors in measuring the fluxes, propagated through
Equation \ref{eq:cext}.

We plotted three theoretical extinction laws in our figures. The WD $R_V = 3.1$
and WD $R_V = 5.5$ models are the same WD3.1 and WD5.5 models we used in
computing the \chisq. In addition to those two, we plotted a third model,
labeled KP v5.0 (Pontoppidian et al., in prep). This model is one from a grid of
models constructed starting from the \citet{weingartner01} parameterization of
the  grain size distribution.  Icy mantles of water and other volatiles were
then added.  The specific model we use from this grid is the one with the ``best
fit'' to the c2d mid-infrared extinction law and ice features.  Several ice
absorption features in this model can be seen in the figures, these are due to
H$_2$O, CO$_2$, or CO.

\subsubsection{ The Extinction Law From $3.6-8\:\mu$m }

In the IRAC bands, the observed extinction laws in all three clouds show a
similar trend: At low \ak, the extinction law is more consistent with the WD3.1
law while at higher extinctions the observed law gradually flattens to become
more consistent with WD5.5. In Table \ref{tab:extlaw} we list our average
extinction law, made from all three of our clouds, along with the results from
other authors and also our three dust models.

In the lowest extinction bin, $0 < A_{K_s} \le 0.5$, the $3.6-8\:\mu$m
extinction law coincides with the law predicted by the WD3.1 dust model, except
for $A_{5.8}/A_{K_s} = 0.28$, which is much higher than the WD3.1 value of 0.17.
The data from Perseus and Serpens are the cause of this; both have an
anomalously high $A_{5.8}/A_{K_s}$ value. If we compute the \chisq{} statistic
and the associated probability, then these data have a $94\%$ chance of  fitting
the WD3.1 model ($\chi^2 = 0.39$), but only a $22\%$ chance of fitting the
WD5.5 model ($\chi^2 = 4.5$).

The extinction law in the $0.5 < A_{K_s} \le 1$ bin appears to be a transition
between the WD3.1 and WD5.5 models.  The WD3.1 $\chi^2$ is $6.8$, only an $8\%$
probability of the data matching the model, while the WD5.5  $\chi^2$ is $3.8$
($29\%$ probability).  For $1 < A_{K_s} \le 2$, our
extinction law is very close to the WD5.5 dust model.  It is extremely unlikely
that the data fit the WD3.1 model here ($\chi^2 = 80$, $2.5\times 10^{-15}\%$
probability). Compared to this, the WD5.5 model has $\chi^2 =1.2$ with a $76\%$
probability. Lastly, for $A_{K_s} > 2$, our average extinction law is slightly
flatter than WD5.5.  This is due to the data points from Perseus and Serpens,
and is  primarily because of the $5.8\:\mu$m data point. The presence of water
ice may explain the $5.8\:\mu$m relative extinction. The KP v5.0 dust model
incorporates the $6.02\:\mu$m water ice absorption line \citep{gibb04}, which is
within the IRAC $5.8\:\mu$m bandpass.  The WD3.1 model is even more strongly
rejected in this bin ($\chi^2 = 303$, $1.9\times 10^{-63}\%$ probability),
though the WD5.5 model is also a relatively poor fit ($\chi^2 = 9.3$, $2.5\%$
probability).

As can be seen in Table \ref{tab:extlaw}, our $3.6-8\:\mu$m extinction law for
$A_{K_s} > 1$ is consistent with the results obtained by other authors.
\citet{lutz99} measured hydrogen recombination lines towards the Galactic center
and computed the extinction law from $2.6-19\:\mu$m. \cite{indebetouw05} used
the IRAC instrument on-board \spitzer{} and measured the extinction law along
two lines of sight using data from the GLIMPSE \spitzer{} Legacy Science program
\citep{benjamin03}. \citet{flaherty07} also used \spitzer{} to measure the
extinction law toward five regions. All of these results are in rough agreement;
they find a flat extinction law from $3.6-8\:\mu$m, consistent with the WD5.5
model.

\subsubsection{ \label{sec:extlawmips} The Extinction Law at $24\:\mu$m }

At $24\:\mu$m, the extinction law is generally higher than that predicted by any
of the three dust models. This means more extinction exists for a given column
density, or alternatively, our observed $24\:\mu$m fluxes are fainter than
models predict. The exact value of $A_{24}/A_{K_s}$ ranges from 0.28 to 1.1 and
varies among clouds and extinction bins. The only $A_{24}/A_{K_s}$ value which
is not higher than any dust model prediction is for $A_{K_s} > 2$ in Ophiuchus,
which is only marginally higher than the predicted value from WD5.5. However,
this value is computed from just 2 stars. This general discrepancy is the reason
we excluded the $24\:\mu$m data when creating our \chisq{} maps in
\S\,\ref{sec:chi2clouds}.

The value of $A_{24}/A_{K_s}$ appears to decrease as \ak{} increases. This
behavior is puzzling because if grain growth is occurring in denser regions, as
suggested by the $3.6-8\:\mu$m extinction law, then we would expect regions with
a larger column density to have a higher value of $A_{24}/A_{K_s}$, not a lower
one, since larger grains should have a larger relative extinction. However, we
observe the opposite trend.

A likely explanation for this trend is that the average stellar model we are
using is incorrect at $24\:\mu$m.  We can obtain an upper-limit on the value for
the average stellar model flux at $24\:\mu$m.  There are 23 stars with
$24\:\mu$m fluxes with $0 < A_{K_s} \le 0.5$ in all three clouds.  If these
sources had zero extinction, then the ratio of $24\:\mu$m to $K_s$ band flux for
these stars would provide an estimate of the true average stellar flux. 
However, these stars \emph{do} have small amounts of extinction, which will
increase the $24\:\mu$m to $K_s$ flux ratio, assuming the true relative
extinction at $24\:\mu$m is less than at $K_s$.  Thus, our ratio of $24\:\mu$m
to $K_s$ band flux is an upper-limit on the true average stellar flux.  For the
23 low-extinction stars in our clouds, the flux ratio is $0.014 \pm 0.002$, the
same value listed in Table \ref{tab:starmodel}. Since the average extinction for
these stars 0.3 \ak, the true value will be less than this. As a test, we took
our combined catalogs for all three clouds and used a smaller average model flux
of $0.012$ mJy at $24\:\mu$m.  In our four extinction bins, our relative 
extinction becomes (from low to high \ak): $0.65\pm0.31$, $0.56\pm0.14$,
$0.52\pm0.08$, and $0.28\pm0.12$.  Using a lower value for the $24\:\mu$m
stellar flux almost completely eliminates the trend of $A_{24}/A_{K_s}$
decreasing as \ak{} increases.  The only extinction bin which shows a
significant drop in the relative extinction is for $A_{K_s} > 2$, but there are
only three $24\:\mu$m stars in this bin.  A larger sample of stars is needed to
confirm this low value.

Other authors also found $A_{24}/A_{K_s}$ to be higher than predicted by dust
models.  Their values are in rough agreement with ours, assuming we use the
lower value for the average stellar flux proposed above.  \citet{flaherty07}
were able to measure $A_{24}/A_{K_s}$ for two of their five regions and found
$A_{24}/A_{K_s} = 0.44\pm0.02$ and $0.52\pm0.03$ for Serpens and NGC2068/71,
respectively. However, they caution that their sample size is small.  Even
though \citet{lutz99} did not measure the extinction longward of $19\:\mu$m, if
their flat extinction law was projected  out to $24\:\mu$m, it would also be
$\sim0.5$.

It is clear that the extinction law at $24\:\mu$m is not well understood.  With
some assumptions about the average stellar flux, our findings are consistent
with those of \citet{flaherty07}.  Both their and our results suggest 
$A_{24}/A_{K_s} \sim 0.5$, or $2-3\times$ larger than the value predicted by
current dust models.  New dust models will need to incorporate additional
extinction at $24\:\mu$m to reproduce these observations.  Furthermore, both 
the Flaherty et al.\ and our results are from relatively small samples.  A
larger study is needed before we can answer questions about variations in the
$24\:\mu$m extinction law with column density or line-of-sight.

\subsubsection{ \label{sec:perseuschi2} Perseus \chisq{} Map }

\begin{figure}

\plotone{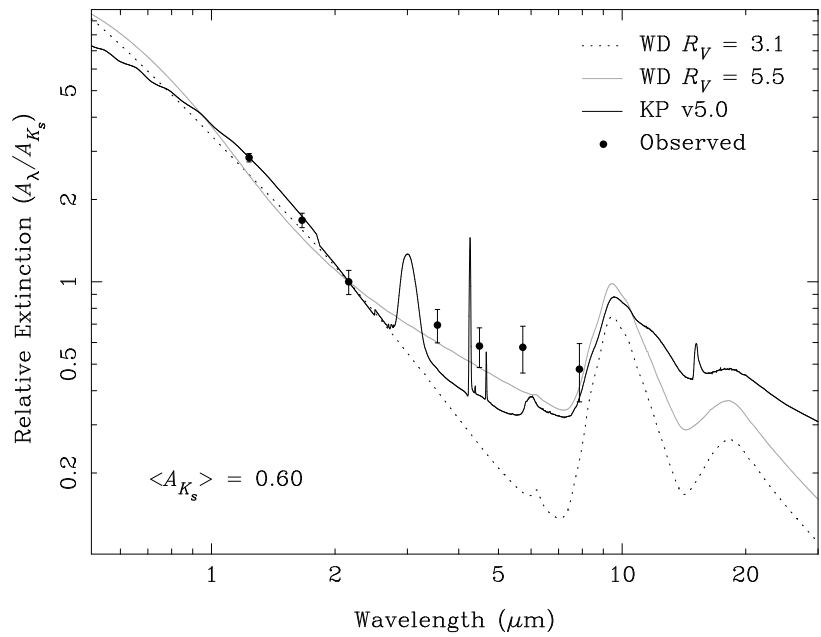}

\caption[ Extinction law of anomalous region within Perseus ]{
\label{fig:perseus-blob} The average extinction law obtained for the region in
Perseus with anomalously high \chisq{} in the $R_V = 3.1$ map (see 
\S\S\,\ref{sec:chi2clouds} and \ref{sec:perseuschi2}). The coordinates of this
region are ($\sim3^\mathrm{h}40.5^\mathrm{m}$ $+31^\circ 30\arcmin$). See Figure
\ref{fig:ophaklaw} for a description of the data points, curves, and errorbars.}

\end{figure}

In this section we will address the `anomalous' region in the Perseus $R_V =
3.1$ \chisq{} map. This region is circled in red in Figure
\ref{fig:perseuschi2}. 53 stars are within the region defined by the $\chi^2 =
4.0$ contour. We plotted the extinction law derived from these 53 stars in
Figure \ref{fig:perseus-blob} and found the extinction law for these data are
significantly higher than the WD5.5 model with the largest deviation occurring
at $5.8\:\mu$m.  The water ice feature at $6.02\:\mu$m may explain the
discrepancy, however ices are seen towards dense cores and the average
extinction for this region is a relatively modest $A_{K_s} = 0.60$
\citep{gibb04,knez05}. Another possibility is that the sources in this region
may not really be stars, but a search of the NASA Extragalactic Database and
Simbad within $2\arcmin$ of $3^{h}40.5^{m}$ $+31^\circ30\arcmin$ yielded no
matches. A final possibility is that the near-infrared extinction law is
different in this region compared to our assumed WD5.5. We will see how this
impacts our mid-infrared extinction law in the next section.

\subsection{ \label{sec:beta-clouds} Changes in the Near-Infrared Extinction 
Law }

As we mentioned in the introduction, the near-infrared extinction law can be fit
by a power law, $A_\lambda \propto \lambda^{-\beta}$, with $\beta=1.6-1.8$
\citep[and references therein]{draine03}. For this paper we assumed the WD5.5
model, with $\beta \approx 1.6$, is correct in the near-infrared. This
assumption allows us to directly compare our results with those of other
authors, who have made the same assumption. Furthermore, our data agree with the
$\beta = 1.6$ reddening vector shown in Figure \ref{fig:ahak-bad}. However, if
the near-infrared extinction law is actually steeper in some regions this will
introduce another source of error into our measurements of the mid-infrared
extinction law. To measure the effect of choosing a different value for $\beta$,
we modified the WD5.5 law to have $\beta=1.8$ in the near-infrared, then
re-classified the stars, computed \ak, and derived the mid-infrared
extinction law. Our values of \alak{} are decreased by using $\beta=1.8$ as
opposed to WD5.5. The decrease is $\sim15\%$ for IRAC1 and IRAC2, $\sim25\%$ for
IRAC3 and IRAC4, and $\sim15\%$ for MIPS1. The extinction law in Figures
\ref{fig:ophaklaw}-\ref{fig:serpensaklaw} still flattens as \ak{} increases, but
does so more gradually. A different value of $\beta$ may partially explain the
`anomalous' region in Perseus. Using the $\beta=1.8$ law for this region, we
found that the derived extinction law in the mid-infrared is decreased to the
point that it is consistent with WD5.5. However, the extinction law is still
flatter than what is observed elsewhere in Perseus at the same \ak{}.

Based on this analysis, we expect our assumption of the WD5.5 extinction law
in the near-infrared may introduce an error of up to $15\%$ in IRAC1, IRAC2, and
MIPS1, and $25\%$ in IRAC3 and IRAC4.  However, we have no way of measuring
possible variations in the near-infrared extinction law along different lines
of sight.  Therefore, while it is important to keep this in mind as a 
\emph{possible} source of error, we have not attempted to incorporate it into
the errorbars shown in our figures.

\section{\label{sec:cloud-discuss} Conclusions }

In this paper we took deep \jhks{} observations of regions within three
molecular clouds: Ophiuchus, Perseus, and Serpens. These were combined with
\spitzer{} c2d data to create band-merged catalogs in eight wavebands from
$1.2-24\:\mu$m. Our final catalogs contain 2,365 stars in Ophiuchus, 11,280 in
Perseus, and 49,485 in Serpens. From these catalogs we created uniform maps of
the extinction and \chisq. We also computed the average extinction law in four
extinction bins: $0 < A_{K_s} \le 0.5$, $0.5< A_{K_s} \le 1$, $1 < A_{K_s} \le
2$, and $A_{K_s} > 2$.

Our results in the IRAC bands, $3.6-8\:\mu$m, show that grain growth occurs in
the dense regions of our clouds. The \chisq{} maps show the deviations from two
dust models: \citet{weingartner01} with $R_V = 3.1$ and $R_V = 5.5$ (abbreviated
as WD3.1 and WD5.5). There is a strong correlation between high extinction and
high WD3.1 \chisq{}, but not with WD5.5 \chisq. This suggests denser regions are
more consistent with the WD5.5 extinction model, which has grain sizes up to
$10\times$ larger than those in the WD3.1 dust model. The same behavior is seen
in the average extinction law. For $0 < A_{K_s} \le 0.5$, it is most consistent
with WD3.1 ($94\%$ probability of the data fitting this model).  As \ak{}
increases, the extinction law gradually flattens and becomes more consistent 
with WD5.5 for $1 < A_{K_s} \le 2$, with a $76\%$ probability.  For $A_{K_s} >
2$, the WD5.5 model is not as good a fit to the data, primarily due to the
$A_{5.8}/A_{K_s}$ relative extinction, which is higher than the model.  This may
indicate the presence of water ice in the densest regions.

Not all areas in our clouds can be explained using grain growth. Our \chisq{}
map of Perseus reveals a region that has only a modest average extinction of
$A_{K_s} = 0.60$, yet has a peak WD3.1 \chisq{} of 39. The extinction law in
this region is much flatter than WD5.5. Because of the low average extinction 
here, it doesn't seem likely that grain growth is altering the extinction
law, nor that ices have formed on the dust grains. It is possible the background
sources in this region are not really stars, or that the near-infrared
extinction law is steeper in this region. More work is needed to understand
the physical cause of the extinction law in this region.

Unlike in the IRAC bands, the $24\:\mu$m extinction law is $2-4\times$ higher
than current dust models predict. There is also a negative correlation between
\ak{} and the value of $A_{24}/A_{K_s}$. Both of these results maybe be
partially mitigated by considering the small numbers of stars detected at
$24\:\mu$m or slight changes to our average stellar model flux. However, even
with the most conservative assumptions, our average value for $A_{24}/A_{K_s}$
is still $\sim2\times$ larger than that predicted by dust grain models. This
results agrees with the measured values from \citet{flaherty07}. Future dust
models will need to reproduce the relatively  large value of the $24\:\mu$m
extinction.

\acknowledgements

We would like to thank the anonymous referee whose useful comments improved
this paper.

Support for this work, utilizing data from the ``Cores to Disks'' Spitzer Legacy
Science Program \citep{evans03}, was provided by NASA through contracts 1224608,
1230782, and 1230779 issued by the Jet Propulsion Laboratory, California
Institute of Technology, under NASA contract 1407. Additional support for
N.~L.C.\ was provided by NASA through JPL contracts 1264793 and 1264492.
L.~G.~M.\ was supported by NASA Origins Grant NAG510611.

This research has made use of PyRAF, a product of the Space Telescope Science
Institute, which is operated by AURA for NASA. We also used SIMBAD and the
NASA/IPAC Extragalactic Database (NED). The SIMBAD database is operated at CDS,
Strasbourg, France. NED is operated by the Jet Propulsion Laboratory, California
Institute of Technology, under contract with the National Aeronautics and Space
Administration.

\appendix
\section{ \label{sec:avgmodel} Average Stellar Models }

To compute the expected stellar distribution for each cloud, we used the galaxy
model from \citet{jarrett92} and \citet{jarrett94}. Given the Galactic
coordinates, distance, and average extinction for a cloud, this model produces
source counts broken down by spectral type. These source counts provide a
rudimentary weighting function that could be used in computing the average
stellar model. However, a weighting function from the raw source counts would
ignore the reality that not all spectral types are equally observable at each
wavelength. Our next step was to correct for the detection limits of our actual
observations. The galaxy model lists the extincted $K_s$ magnitude for each star
and its extinction, $A_V$. We used this information, the Weingartner \& Draine
$R_V = 3.1$ extinction law, and the stellar templates from
``star-pet''\footnote{\url{http://ssc.spitzer.caltech.edu/tools/starpet}} to
compute the magnitude of each star at every observed waveband from $J-24\:\mu$m.
We used WD3.1 since this was the law used by the galaxy model when computing the
extinctions. We arbitrarily discarded any detection falling below the $5\sigma$
cutoff for the given waveband (Table \ref{tab:sigmalimits}). Using the
normalized, final, corrected source counts at each waveband as a weighting
function, we then computed the average stellar template for each region and an
overall average stellar model. Our results are shown in Table
\ref{tab:starmodel}. We repeated the above procedure using the Weingartner \&
Draine $R_V = 5.5$ extinction law. The average fluxes using this law never vary
by more than \mbox{0.001 mJy} from those listed in Table \ref{tab:starmodel}.

\clearpage

\begin{deluxetable}{lcccc}
\tablewidth{0pt}

\tablecolumns{5}

\tablecaption{\label{tab:cloudbasic} Basic Properties of the Clouds }

\tablehead{\colhead{} & \colhead{$l$} & \colhead{$b$} & \colhead{Dist.} 
& \colhead{ Area Mapped in \jhks{} } \\

\colhead{ Cloud } & \colhead{(deg.)} & \colhead{(deg.)} & \colhead{(pc)} 
& \colhead{ (deg$^2$)}}

\startdata
Ophiuchus  & 353 &  16 & $125\pm25$\tablenotemark{a} & 0.15\\
Perseus    & 160 & -19 & $250\pm50$\tablenotemark{b} & 1.0\\
Serpens    &  30 &   5 & $260\pm10$\tablenotemark{c} & 0.33\\
\enddata

\tablenotetext{a}{\citet{degeus89}}
\tablenotetext{b}{\citet{enoch06}}
\tablenotetext{c}{\citet{straizys96}}
\end{deluxetable}

\begin{deluxetable}{ccc}
\tablewidth{0pt}
\tablecolumns{3}
\tablecaption{ \label{tab:sigmalimits} Magnitude Limits}
\tablehead{\colhead{Band} & \colhead{$10\sigma$ Limits} & 
\colhead{$5\sigma$ Limits}}
\startdata
$J$   & 19.5 & 20.3\\
$H$   & 18.8 & 19.5\\
$K_s$ & 17.7 & 18.4\\
IRAC1 & 16.6 & 17.3\\
IRAC2 & 15.9 & 16.6\\
IRAC3 & 13.6 & 14.3\\
IRAC4 & 12.8 & 13.5\\
MIPS1 &  7.7 &  8.4\\
\enddata
\end{deluxetable}

\begin{deluxetable}{lccccc}
\rotate
\tablewidth{0pt}
\tablecolumns{6}
\tablecaption{ \label{tab:extlaw} Average Relative Extinction \alak}
\tablehead{\colhead{Source} & \colhead{$3.6\:\mu$m} & \colhead{$4.5\:\mu$m} &
\colhead{$5.8\:\mu$m} & \colhead{$8\:\mu$m} & \colhead{$24\:\mu$m}}

\startdata
Our data, $0 < A_{K_s} \le 0.5$ & $0.41 \pm 0.19$ & $0.26 \pm 0.18$ &
$0.28 \pm 0.18$ & $0.21 \pm 0.17$ & $1.08 \pm 0.32$\\
Our data, $0.5 < A_{K_s} \le 1$ & $0.49 \pm 0.10$ & $0.35 \pm 0.10$ &
$0.35 \pm 0.10$ & $0.35 \pm 0.10$ & $0.75 \pm 0.14$\\
Our data, $1 < A_{K_s} \le 2$   & $0.60 \pm 0.05$ & $0.46 \pm 0.05$ &
$0.44 \pm 0.05$ & $0.43 \pm 0.05$ & $0.61 \pm 0.08$\\
Our data, $A_{K_s} \ge 2$       & $0.64 \pm 0.03$ & $0.53 \pm 0.03$ &
$0.46 \pm 0.03$ & $0.45 \pm 0.03$ & $0.34 \pm 0.13$\\
\cutinhead{Other Authors}
\citet{flaherty07}   & $0.632 \pm 0.004$ & $0.54 \pm 0.01$ & 
$0.50 \pm 0.02$ & $0.50 \pm 0.01$ & $0.46 \pm 0.04$\\
\citet{indebetouw05} & $0.56 \pm 0.06$   & $0.43 \pm 0.08$ &
$0.43 \pm 0.10$ & $0.43 \pm 0.10$ & \nodata\\
\citet{lutz99}\tablenotemark{a}&$0.53 \pm 0.03$   & $0.50 \pm 0.08$ &
$0.49 \pm 0.06$ & $0.42 \pm 0.06$ & \nodata\\
\cutinhead{Dust Models\tablenotemark{b}}
\citet{weingartner01}, $R_V = 3.1$ & 0.40 & 0.25 & 0.17 & 0.22 & 0.17\\
\citet{weingartner01}, $R_V = 5.5$ & 0.60 & 0.49 & 0.40 & 0.41 & 0.24\\
KP, v5.0 & 0.48 & 0.38 & 0.34 & 0.38 & 0.38\\
\enddata
\tablenotetext{a}{The extinction for the closest ISO wavelength to each 
\spitzer{} band is listed: 3.7, 4.4, 5.9, and $7.5\:\mu$m}
\tablenotetext{b}{Extinctions are computed at the central wavelength in each
band}

\end{deluxetable}

\begin{deluxetable}{lcccccccc}
\rotate
\tablewidth{0pt}
\tablecolumns{9}
\tablecaption{ \label{tab:starmodel} Average Stellar Model for Each Cloud }
\tablehead{ \colhead{ Region } & \colhead{$J$} & \colhead{$H$} & 
\colhead{$K_s$} & \colhead{IRAC1} & \colhead{IRAC2} & \colhead{IRAC3} &
\colhead{IRAC4} & \colhead{MIPS1}\\
\colhead{} & \colhead{(mJy)} & \colhead{(mJy)} & \colhead{(mJy)} & 
\colhead{(mJy)} & \colhead{(mJy)} & \colhead{(mJy)} & \colhead{(mJy)} &
\colhead{(mJy)}}
\startdata

Ophiuchus & $1.374\pm0.291$ & $1.335\pm0.095$ & 1. & $0.448\pm0.018$ & 
   $0.267\pm0.011$ & $0.178\pm0.008$ & $0.103\pm0.006$ & $0.014\pm0.001$\\
Perseus   & $1.238\pm0.276$ & $1.284\pm0.099$ & 1. & $0.457\pm0.023$ &
   $0.269\pm0.011$ & $0.180\pm0.013$ & $0.101\pm0.008$ & $0.014\pm0.000$\\
Serpens   & $1.560\pm0.232$ & $1.393\pm0.061$ & 1. & $0.443\pm0.013$ & 
   $0.261\pm0.016$ & $0.177\pm0.005$ & $0.106\pm0.005$ & $0.014\pm0.000$\\
\hline
Average   & $1.412\pm0.152$ & $1.357\pm0.046$ & 1. & $0.447\pm0.010$ &
   $0.267\pm0.007$ & $0.178\pm0.004$ & $0.104\pm0.003$ & $0.014\pm0.001$\\
\enddata
\end{deluxetable}

\clearpage

\clearpage

\end{document}